\documentclass[conference]{IEEEtran}
\usepackage{nicefrac}
\usepackage{siunitx}
\usepackage{array,framed}%array用来修改表格行高
\usepackage{booktabs}
\usepackage{
  color,
  float,
  epsfig,
  wrapfig,
  graphics,
  graphicx,
  subcaption
}
\usepackage{textcomp,amssymb}
\usepackage{setspace}
\usepackage{latexsym,fancyhdr,url}
\usepackage{enumerate}
\usepackage{algorithm2e}
\usepackage{algpseudocode}
\usepackage{graphics}
\usepackage{xparse} % argument parsing -- \edist
\usepackage{xspace}
\usepackage{multirow}
\usepackage{csvsimple}
\usepackage{balance}
\usepackage{amsmath}
\usepackage{longtable}
\usepackage{supertabular}
\usepackage[table]{xcolor}
\usepackage{colortbl}
\usepackage{multicol}
\usepackage{afterpage}

\usepackage{
  tikz,
  pgfplots,
  pgfplotstable
}
\usepackage{hyperref}
\usepackage{ulem}

\usetikzlibrary{
  shapes.geometric,
  arrows,
  external,
  pgfplots.groupplots,
  matrix
}

\newcommand\ignore[1]{}
\newcommand\delete[1]{}
\usepackage{xparse}
\usepackage{listings}
\newcommand{\bnm}{\begin{newmath}}
\newcommand{\enm}{\end{newmath}}

\newcommand{\bea}{\begin{eqnarray*}}%
\newcommand{\eea}{\end{eqnarray*}}%

\newcommand{\bne}{\begin{newequation}}
\newcommand{\ene}{\end{newequation}}

\newcommand{\bal}{\begin{newalign}}
\newcommand{\eal}{\end{newalign}}

\newenvironment{newalign}{\begin{align}%
\setlength{\abovedisplayskip}{4pt}%
\setlength{\belowdisplayskip}{4pt}%
\setlength{\abovedisplayshortskip}{6pt}%
\setlength{\belowdisplayshortskip}{6pt} }{\end{align}}

\newenvironment{newmath}{\begin{displaymath}%
\setlength{\abovedisplayskip}{4pt}%
\setlength{\belowdisplayskip}{4pt}%
\setlength{\abovedisplayshortskip}{6pt}%
\setlength{\belowdisplayshortskip}{6pt} }{\end{displaymath}}

\newenvironment{newequation}{\begin{equation}%
\setlength{\abovedisplayskip}{4pt}%
\setlength{\belowdisplayskip}{4pt}%
\setlength{\abovedisplayshortskip}{6pt}%
\setlength{\belowdisplayshortskip}{6pt} }{\end{equation}}

\newcounter{ctr}

\newcounter{mytable}
\def\mytable{\begin{centering}\refstepcounter{mytable}}
\def\endmytable{\end{centering}}

\newcounter{myfig}
\def\myfig{\begin{centering}\refstepcounter{myfig}}
\def\endmyfig{\end{centering}}

\newlength{\saveparindent}
\newlength{\saveparskip}
\newcommand{\E}{{\rm I\kern-.3em E}}

\renewcommand{\eqref}[1]{\mbox{Equation~(\ref{#1})}}

\def \part {part}

\renewcommand{\paragraph}[1]{\vspace*{6pt}\noindent\textbf{#1}\;}

\def \blackslug{\hbox{\hskip 1pt \vrule width 4pt height 8pt
    depth 1.5pt \hskip 1pt}}
\def \qed{\quad\blackslug\lower 8.5pt\null\par}
\newcounter{mynote}[section]

\newcounter{rcnote}[section]

\newcounter{mrnote}[section]

\newcounter{fknote}[section]

\newcounter{anote}[section]

\DeclareMathSymbol{\mlq}{\mathord}{operators}{``}
\DeclareMathSymbol{\mrq}{\mathord}{operators}{`'}

\newcommand{\rhf}[2]{R_{f, \gamma}}

\DeclareDocumentCommand{\edist}{o o}{
  \ensuremath{
    \IfNoValueTF{#1}{{d}}{{\sf d}(#1,#2)}
  }
}

\newcommand{\olrk}[1]{\ifx\nursymbol#1\else\!\!\mskip4.5mu plus 0.5mu\left(\mskip0.5mu plus0.5mu #1\mskip1.5mu plus0.5mu \right)\fi}

\NewDocumentCommand{\indseq}{ O{1} O{r} }{{#1}\ldots {#2}}

\usepackage{tikz}

\newcommand{\tabincell}[2]{\begin{tabular}{@{}#1@{}}#2\end{tabular}}%表格自动换行

\usepackage{threeparttable}

\usepackage{pifont}
\definecolor{darkgreen}{HTML}{24A10B}
\newcommand {\toolname}[1]{GPTAid}
\newcommand {\ablationRawRecall}[1]{84.4\%}
\newcommand {\ablationRawPre}[1]{11.9\%}
\newcommand {\ablationFilterPre}[1]{79.1\%}
\newcommand {\ablationFinalPre}[1]{92.3\%}
\newcommand {\ablationFinalRecall}[1]{71.0\%}
\newcommand {\rightSuccessRate}[1]{93.5\%}
\newcommand {\confirmedAPSR}[1]{76}
\newcommand{\dels}[1]{}
\usepackage{tikz}
\newcommand{\todo}[1]{{\textcolor{red}{[TODO: #1]}}}
\newcommand{\yy}[1]{{\textcolor{black}{#1}}}
\newcommand{\jh}[1]{{\textcolor{black}{#1}}}
\newcommand{\change}[1]{{\textcolor{black}{#1}}}
\newcommand{\jht}[1]{{\textcolor{black}{#1}}}
\newcommand\hit[1]{}
\ifCLASSINFOpdf
\else
\fi
\begin{document}
%
% paper title
% can use linebreaks \\ within to get better formatting as desired
\title{Generating API Parameter Security Rules with LLM for API Misuse Detection}

% author names and affiliations
% use a multiple column layout for up to three different
% affiliations
% \author{
% \IEEEauthorblockN{Jinghua Liu$^{1,2}$, Yi Yang$^{1,2,\star}$, Kai Chen$^{1,2,\star}$\thanks{\star\ \text{Corresponding Author}}, and Miaoqian Lin$^{1,2}$}\\
% \IEEEauthorblockA{
% $^1$Institute of Information Engineering, Chinese Academy of Sciences, China\\
% $^2$School of Cyber Security, University of Chinese Academy of Sciences, China\\
% \email{\{liujinghua, yangyi, chenkai, linmiaoqian\}@iie.ac.cn}
%     }
% }

\author{
\IEEEauthorblockN{Jinghua Liu$^{1,2}$, Yi Yang$^{1,2,\star}$, Kai Chen$^{1,2,\star}$\thanks{$\star$\ \text{Corresponding Author}}, and Miaoqian Lin$^{1,2}$}\\
\IEEEauthorblockA{
$^1$Institute of Information Engineering, Chinese Academy of Sciences, China\\
$^2$School of Cyber Security, University of Chinese Academy of Sciences, China\\
{\{liujinghua, yangyi, chenkai, linmiaoqian\}@iie.ac.cn}
    }
}

\IEEEoverridecommandlockouts
\makeatletter\def\@IEEEpubidpullup{6.5\baselineskip}\makeatother
\IEEEpubid{\parbox{\columnwidth}{
		Network and Distributed System Security (NDSS) Symposium 2025\\
		23-28 February 2025, San Diego, CA, USA\\
		ISBN 979-8-9894372-8-3\\
		https://dx.doi.org/10.14722/ndss.2025.23465\\
		www.ndss-symposium.org
}
\hspace{\columnsep}\makebox[\columnwidth]{}}

% make the title area
\maketitle
\fancyhf{} 
\fancyfoot[C]{\thepage}

% IEEEtran.cls defaults to using nonbold math in the Abstract.
% This preserves the distinction between vectors and scalars. However,
% if the conference you are submitting to favors bold math in the abstract,
% then you can use LaTeX's standard command \boldmath at the very start
% of the abstract to achieve this. Many IEEE journals/conferences frown on
% math in the abstract anyway.

% no keywords

% For peer review papers, you can put extra information on the cover
% page as needed:
% \ifCLASSOPTIONpeerreview
% \begin{center} \bfseries EDICS Category: 3-BBND \end{center}
% \fi
%
% For peerreview papers, this IEEEtran command inserts a page break and
% creates the second title. It will be ignored for other modes.
%%\IEEEpeerreviewmaketitle

% Section I
\begin{abstract}
When utilizing library APIs, developers should follow the API security rules to mitigate the risk of API misuse. \jht{API Parameter Security Rule (APSR) is a common type of security rule that specifies how API parameters should be safely used and places constraints on their values.}
Failure to comply with the APSRs can lead to severe security issues, including null pointer dereference and memory corruption.
\jht{Manually analyzing numerous APIs and their parameters to construct APSRs is labor-intensive and needs to be automated.}
% 已有工作从doc和code中提取信息，然而由于信息的缺失以及分析算法的不全面，导致规则存在缺失。
\jht{
Existing studies generate APSRs from documentation and code, but the missing information and limited analysis heuristics result in missing APSRs.
}
% \jht{
% Existing studies face two main challenges: (i)They generate APSRs from documentation and code, but the lack of comprehensive information and the tools' inability to fully explore the data result in missing APSRs.
% (ii)They struggle to differentiate between security-critical and functionally relevant rules.
% }
% A promising way to generate APSRs is analyzing API source code using static analysis techniques with predefined rules. 
% API源码与API的operation是一致的，
Due to the superior Large Language Model's (LLM) capability in code analysis and text generation without predefined heuristics, we attempt to utilize it to address the challenge encountered in \yy{API misuse detection}.
% 过于宽泛的规则难以被转换成正确具体的检测代码来检测bug，可能导致检测出错误的bug以及漏掉bug
%\jht{However, directly utilizing LLMs leads to incorrect (false bugs) and overly general APSRs. These general APSRs make it challenging to generate correct, concrete detection code, leading to false bugs and missed bugs. Addressing these issues is urgent for bug detection.}
\yy{
However, directly utilizing LLMs leads to incorrect APSRs which may lead to false bugs in detection, and overly general APSRs that could not generate applicable detection code resulting in many security bugs undiscovered. 
}

In this paper, we present a new framework, named \toolname{}, for automatic APSRs generation by analyzing API source code with LLM and detecting API misuse \change{caused by incorrect parameter use}. 
To validate the correctness of the LLM-generated APSRs, we propose an execution feedback-checking approach based on the observation that security-critical API misuse is often caused by APSRs violations, and most of them result in runtime errors.
\ignore{Specifically, \toolname{} first generates raw APSRs and the Right calling code based on the API source code analysis.
%and the Right calling code through a step-by-step promtping together with automatic program repair with LLM's assistance for correctness validation.
Violation codes are then generated for each raw APSRs based on the Right calling code modification with LLM, 
\toolname{} performs dynamic execution for each piece of violation code are further filters out the APSRs based on runtime errors. }
\jht{
Specifically, \toolname{} first uses LLM to generate raw APSRs and the Right calling code, and then generates Violation code for each raw APSR by modifying the Right calling code using LLM. Subsequently, \toolname{} performs dynamic execution on each piece of Violation code and further filters out the incorrect APSRs based on runtime errors.}
To further generate concrete APSRs, \toolname{} employs a code differential analysis to refine the filtered ones.
Particularly, as the programming language is more precise than natural language, \toolname{} identifies the key operations within Violation code by differential analysis, and then generates the corresponding concrete APSR based on the aforementioned operations.
These concrete APSRs could be precisely interpreted into applicable detection code, which proven to be effective in API misuse detection.
%We propose an approach to refine APSRs into a concrete form using code differential analysis, leveraging the precision of formal programming languages (code) over natural language (text). 
%Specifically, \toolname{} identifies the key operations which violates APSR within the code by analyzing the differences, and then generates the corresponding concrete APSR based on these operation codes.
%\jht{To detect API misuse}, we utilize CodeQL together with these concrete APSRs for API misuse detection.
% \todo{Implementing on the dataset containing 200 randomly selected APIs from eight popular libraries, \toolname{} generated \change{6} times more APSRs than the state-of-the-art detectors with a precision of \ablationFinalPre{}. }
Implementing on the dataset containing 200 randomly selected APIs from eight popular libraries, \toolname{} achieves a precision of \ablationFinalPre{}. Moreover, it generates 6 times more APSRs than state-of-the-art detectors on a comparison dataset of previously reported bugs and APSRs.
We further evaluated \toolname{} on 47 applications, \change{210} unknown security bugs were found potentially resulting in severe security issues (e.g., system crashes), \change{150} of which have been confirmed by developers after our reports.
%\todo{abstract needs to be revised}

% %%%%%%%%%%%%IGNORE%%%%%%%%%%%%%%%%%%%%%%%%%%%%%%%%%%

\end{abstract}

\section{Introduction}
\label{sec:intro}
%%%%%%%%%%%%%%%%%%%%%%%%%%%%%%%
\begin{figure}
    \centering
    \includegraphics[width=1\linewidth]{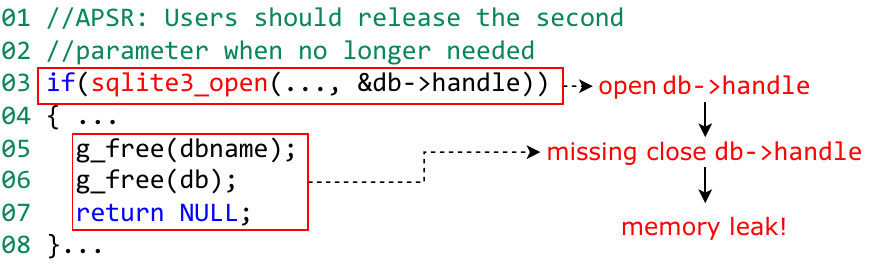}
    \caption{Example for an API misuse in darktable} 
\label{fig:intro-APSR-example}
\vspace{-5pt}
\end{figure}
%%%%%%%%%%%%%%%%%%%%%%%%%%%%%%%
Today, Application Programming Interfaces (APIs) play a vital role in software development, which enables developers to reuse functions from software libraries.
% The information asymmetry between API developers and API users may result in unintentionally API misuses, leading to significant security issues.
% 用户在使用API时需要遵守一些参数限制，返回值限制以及invocation context限制
API security rules should be strictly followed by software developers and could be classified as parameter rules (e.g., \textit{``parameter must not be NULL''}), rules focusing on return value (e.g., \textit{``return value must be checked against NULL''}) and the rules involving invocation condition (e.g., \textit{``must be called before any other action takes place''}).
Violating security rules can result in significant security issues, such as memory corruption, Denial-of-Service, and so on. 
API parameter security rules (APSRs) is one of the common types of security rules that have been extensively studied in the previous research~\cite{ZhouAnalysing2017,Zhongempir2020}. Specifically, APSRs specify the security rules on parameter values (e.g., \textit{``parameters must not be negative''}) and the parameter-associated operations (e.g., \textit{``must not be freed''}). 
% Based on our analysis of known API misuse, XX\% of them result from violating of APSRs.
According to the thorough analysis of 100 randomly selected known misuses from the existing work~\cite{jiang2024appminer,Lyugoshawk2022,lin2023aphp,lv2020rtfm}, we found 71\% of them resulted from APSR violations.
% 违反APSR会导致严重的安全问题，比如memory leak， 空指针解引用，buffer overflow等
For example, Figure~\ref{fig:intro-APSR-example} illustrates the misuse of the API \texttt{sqlite3\_open} from the \textit{darktable} application. 
One APSR of \texttt{sqlite3\_open} says: \textit{``release the second parameter when no longer needed''.} 
According to the figure, the second parameter \texttt{db->handle} is allocated in Line 3, however, the caller fails to release the allocated resource in case \texttt{sqlite3\_open} fails within Line 5 to Line 7, which violates this APSR and leads to a memory leak.
%Without APSRs, analyzers (e.g., CodeQL) lack the necessary context to identify API misuse, as these APSRs provide the specific constraints that define safe API usage.
With the extensive information from APSRs, automatic analyzers can detect API misuses and support secure software development, making automatic APSR generation essential.

% %%%%%%%%%%%%%%%%已有工作%%%%%%%%%%%%%%%%%%%%%%%%
% API误用的原因是违反了对应的APSR

% Due to the lack of APSR, detecting parameter-related API misuse is challenging.
% 已有的方法用来：
% 已有生成APSR和检测API误用的方法大多从document提取信息生成APSR或者对比API调用代码（client code）中的多个usage pattern来检测与大多数pattern不同的API调用作为API误用。
% 然而，由于文档中对于APSR描述的缺失，利用文档信息生成APSR会造成大量的missing。利用client code检测API误用会导致错误的结果due to the correct usage pattern is hard to determine.
Prior studies on APSRs generation and API misuse detection typically involve several limitations. Documentation-based approaches~\cite{lv2020rtfm, ren2020api, ZhouAnalysing2017, Zhongempir2020, huaurc2023} generate APSRs based on the extracted knowledge, which may lead to security bugs undetected due to the absence of APSR description.
The approaches based on the API calling code~\cite{kang2016apex, yun2016apisan,liuIPPO2021,Wenexpose2019, huaurc2023} generate APSRs by comparing usage patterns, which may also yield false negatives (Section~\ref{sec:compare-study}) resulting from the difficulties in identifying the correct patterns.
%And the APSRs generated by execution traces analysis~\cite{le2018deepm, kang2021specm} merely reveal the coarse-grained call order, that lacks value constraints and results in \jht{incomplete APSRs for misuse detection.}
%Finding a suitable data source for accurate APSRs generation is one of the prerequisite conditions that must be addressed.
%Usage patterns comparison within the API calling code to detect deviations as potential misuses and specifications mining based on execution traces of test cases~\cite{le2018deepm, kang2021specm}. However, documentation-based methods cause a lot of security bugs undetected due to the absence of APSR descriptions; 
%the calling-code methods also yield ;
%Additionally, given the difficulty in identifying the correct patterns, the analysis of API calling code for API detection also yields false negatives (Section~\ref{sec:compare-study}). 
% 在不知道API内部细节的情况下，生成的test case难以覆盖全面，而且通过execution trace挖掘出来的specification只能体现粗粒度的调用次序的规则，而不能挖掘出值约束
%Furthermore, lack of API knowledge hinders the generation of comprehensive and targeted test cases, and specifications mined from execution traces reveal only coarse-grained call order, missing value constraints.
%\todo{add one more sentence to conclude}
% 我们发现API的source code是一个生成APSR的好的信息来源。API的源码与API的functionality是一致的。通过分析API源码得到的APSR可以被用来检测parameter related的API误用。我们注意到存在一些工作通过分析API源码检测API误用。然而，它们limited to fixed types of bugs 因为它们使用预定义的源码分析方式以分析源码中的某些statement（）
Different from the former types of data, API source code is found to be a solid resource as it is the actual implementation of the API's functionality. 
Existing studies have shown its effectiveness~\cite{huaurc2023, Lyugoshawk2022, ZhouAnalysing2017} on API misuse detection. 
However, they are limited to specific bug types, 
%such as memory management issues, 
due to the high reliance on predefined rules which are designed specifically for certain code statements.
%(such as  \texttt{malloc()}). }
%\jht{However, these studies are limited to identifying specific types of bugs, such as memory management issues, because they depend on predefined static analysis rules. These rules are defined to detect certain statements within the API source code (like \texttt{malloc()}), which may not cover other types of bugs.}
Unfortunately, the analysis of various statements within the API source code is time-consuming and needs complex data/control flow analysis and natural language rules construction for each piece of analysis.
With the superior capability of Large Language Models (LLMs) on code analysis and text generation, enabling the analysis of both programming and natural languages at the same time, we propose to exploit LLMs for APSRs generation. 
\noindent\textbf{Challenges in using LLMs.} 
However, directly using LLM for APSRs generation is challenging.
\uline{\textbf{C1: Incorrect APSRs.}}
The first challenge is the incorrectly generated APSRs by LLMs possibly due to hallucination~\cite{li2023hallucination}. For example, when prompting one of the LLMs with the Prompt \textit{``When using the standard library function ``\texttt{free}'' in C, what are the API parameter security rules the caller needs to follow to prevent security issues?''}, the LLM's output says: \textit{``Before calling \texttt{free}, always check if the pointer is NULL.''}. However, this response is incorrect because the \texttt{free} function does not pose a security risk when its argument is NULL. 
\ignore{A deeper}Our analysis shows that the accuracy of directly using LLM for APSRs generation is merely \ablationRawPre{} (Section ~\ref{sec:ablation}), which needs to be addressed for precisely APSRs generation.
What's more, verifying correct APSRs also poses difficulties, which may result from missing correct references for comparison and preventing the existing evaluation methods (e.g., BLEU~\cite{papineni-etal-2002-bleu}).
Therefore, how to generate correct APSRs including the verification remains the first challenge in need of addressing.
\begin{figure}
    \centering
    \includegraphics[width=1\linewidth]{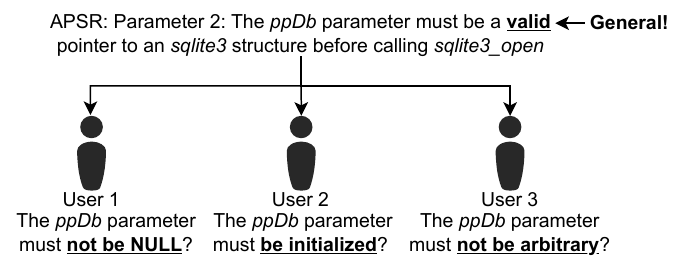}
    \caption{Overly general APSR leads to incorrect interpretations} 
    \vspace{-5pt}
\label{fig:ambiguity-example}
\end{figure}
%%%%%%%%%%%%%%%%%%%%%
% 第二个挑战是LLM生成的结果过于general可能导致歧义。比如图XX中展示了一个general的APSR对于不同的人来说可能有着不同的理解。图中的APSR中描述了ppdb必须是valid的。然而对于不同人来说，valid可能指的是not NULL, 可能指的是be initialized before，也可能是不能指向随意构造的地址。在使用APSR时不同的理解可能会导致错误的操作。不幸的是，由于APSR是自然语言，判断自然语言是否是general的也是一个困难的任务。
\uline{\textbf{C2: Overly-general APSRs.}} 
The second challenge is the generated APSRs might be overly general, which is hard to be interpreted into accurate APSRs and applicable detection codes for API misuse detection.
For example, Figure~\ref{fig:ambiguity-example} illustrates one piece of APSR generated by LLM, saying ``\textit{the \texttt{ppdb} must be \textit{valid}}'', which is too general.
Specifically, \textit{``valid''} could be variously interpreted by different users as ``not NULL'', ``requiring initialization'', and ``not being arbitrary values''. However, only the explanation \textit{``the \texttt{ppdb} must not be NULL''} is correct which could be precisely interpreted into applicable detection code.
Existing approaches have shown effectiveness in using Word Sense Disambiguation (WSD)~\cite{navihli2009wsds, ijcai2021wsds2} technique to eliminate natural language's ambiguity, while it fails to generate accurate APSRs due to its difficulties in combination with security knowledge.  
Since incorrectly interpreting the APSRs introduces false positives and false negatives in API misuse detection, proposing an approach to generate accurate APSRs to eliminate the over-generalization is an urgent need.

%The second challenge is the generated APSRs might be overly general, making it challenging to interpret them into accurate and concrete detection code for API misuse detection.For example, as shown in Figure~\ref{fig:ambiguity-example}, the APSR ``\textit{the \texttt{ppdb} must be \textit{valid}}'' generated by LLMs is too general.
% 这个APSR真正指示的限制是ppdb must not be NULL。因此在根据该APSR检测API误用时可能需要检测missing check的问题。
%Specifically, \textit{``valid''} could be variously interpreted by different users as ``not NULL'', ``requiring initialization'', and ``not being arbitrary values''. However, only the explanation \textit{``the \texttt{ppdb} must not be NULL''} is correct which could be precisely interpreted into applicable detection code.
% 将该APSR理解为其他限制会导致检测API误用时根据不同的检测规则进行检测，导致错误的检测结果。
%Incorrectly interpreting the APSRs introduces false positives and false negatives in API misuse detection.
% 已有NLP方法使用WSD技术消除文本中的多义性，然而该技术依赖于训练数据集的规模，而且难以结合安全知识与上下文语境信息消除规则中对于operation描述的不明确，这使生成非歧义的规则非常困难。
%\jht{Existing approaches use Word Sense Disambiguation (WSD)~\cite{navihli2009wsds, ijcai2021wsds2} technique to eliminate ambiguity in text. However, WSD struggles to integrate security knowledge effectively into text analysis, making it hard to identify ambiguous security-related terms. This limitation significantly affects its use in generating concrete APSRs. Therefore, proposing an approach to generate concrete APSRs is urgently needed.}
% Our work:
% evaluation re:
% contribution

%%%%%%%%%%%%%%%%%%%%%%Our WORK %%%%%%%%%%%%%%%%%%%%%%%%%%%%%%%%%%%%%%%%%%%%%%%%%%%%%%%%%%%%
\noindent\textbf{Our work.} 
% Advance
% \todo{xxx}
In this paper, we proposed \toolname{} (short for \underline{G}enerating API \underline{P}arameter securi\underline{T}y rules from \underline{A}PI source code for API m\underline{I}suse \underline{D}etection) -- a tool for automatically generating accurate and concrete APSRs with LLM and detecting API misuse \change{caused by incorrect parameter use}.
\toolname{} addresses the aforementioned challenges based on several observations.
Since the API source code is the actual implementation of APIs, \toolname{} first prompts to generate the raw APSRs based on the API source code.
To verify the correctness of the raw APSRs, \toolname{} adopts an execution feedback-checking approach to validate the LLM-generated APSRs.
More specifically, we observe that violations of APSRs often lead to security-related API misuse and most of them result in runtime errors which can be caught by the monitoring tools (e.g., sanitizer).
For example, Figure~\ref{fig:intro-APSR-example} describes one piece of APSR specifying \textit{``parameter 2 must be released after calling \texttt{sqlite3\_open}''}. 
The code snippet shows a misuse caught by the sanitizer in a real application which results in a memory leak.
Based on this observation, we separate the validation process into three parts: Right code ($C_r$) generation, Violation code ($C_v$) generation and correct APSRs verification.
\ignore{Particularly, \toolname{} proposes to generate the Right calling code based on a step-by-step approach with the API source code and to ensure accuracy through an LLM-empowered automatic program repair method to automatically repair the erroneous code by monitoring the output of compilation and execution with LLMs.}
\jht{Particularly, \toolname{} proposes to generate the ($C_r$) based on a step-by-step approach with the API source code. To ensure accuracy, \toolname{} monitors execution outputs and applies an LLM-empowered automatic program repair method to automatically repair the erroneous code.}
Then, \toolname{} generates the \jht{($C_v$)} which violates the raw APSRs by modifying the \jht{($C_r$)}.
To make sure the \jht{($C_v$)} correctly violates the APSRs, \toolname{} \jht{applies} program repair approach which is also adopted in the previous step.
Besides, a deeper analysis is conducted to filter out the incorrect \jht{($C_v$)}, including applying the off-the-shelf static analyzer to locate API calls and parameters, and further checking the inconsistency between them (Section~\ref{sec:method-validation}).
At last, \toolname{} dynamically executes the \jht{($C_v$)} assisted by sanitizers, which output runtime errors for correct APSRs and success for incorrect APSRs. 
To solve the overly-general APSRs, we propose to adopt code differential analysis to generate concrete APSRs,
\jht{as the programming language, being a formal language, describes the API more specifically than natural language.
}
Take the APSR \textit{``Parameter 2: The ppDb parameter must be a valid pointer to an sqlite3 structure before calling sqlite3\_open,''} as an example.
The calling code that violates this piece of APSR is \texttt{sqlite3\_open(..., NULL)}, which sets the second parameter to \texttt{NULL} and exactly violates the APSR \textit{``...must be a \uline{valid} pointer...''}. 
Contrary to natural language, the violation code merely leads to one interpretation resulting in no confusion for detection code generation from APSRs.
Leveraging this observation, we propose to analyze the modification operations between ($C_r$) and ($C_v$) to generate concrete APSRs.
However, a large amount of redundant information exists in the code, which hinders the modified operation identification and contributes to the arising of inaccurate APSRs. 
To diminish the redundant code, \toolname{} instructs LLM to identify the shared modified operation among different violation code that leads to the same runtime error.
According to that, \toolname{} acquires the concrete APSRs for each key modification that causes the API-related runtime errors with LLM (Section~\ref{sec:method-refine}).

We randomly selected 200 APIs from eight widely used libraries, including OpenSSL~\cite{openssl}, SQLite~\cite{sqlite3doc}, libpcap~\cite{libpcap}, libxml2~\cite{libxml2}, libevent~\cite{libevent}, libzip~\cite{libzip}, zlib~\cite{zlib} and libcurl~\cite{libcurl}, and constructed a new dataset by analyzing documentation and API source code.
We evaluated the effectiveness of \toolname{} on APSRs generation on this dataset and the results show that \toolname{} achieves a precision of \ablationFinalPre{} and a recall of \ablationFinalRecall{}. \toolname{} identifies eight distinct types of APSRs, which surpasses the performance of previous work~\cite{ZhouAnalysing2017, Zhongempir2020} with two more rule categories identified (Section~\ref{sec:apsr_type}).
% 该recall明显超出了已有的sota工具，advance的recall只有5.4， goshawk只有9%
\toolname{} outperforms the existing state-of-the-art tools, such as Advance~\cite{lv2020rtfm}), which generates only one-seventh as many APSRs as \toolname{} due to incomplete documentation (Section~\ref{sec:compare-study}).
% Advance's~\cite{lv2020rtfm} recall is 5.4\%, while Goshawk's~\cite{Lyugoshawk2022} is 9\%.
In total, \toolname{} found \change{210} unknown API misuses from 47 applications integrating on eight libraries, of which \change{150} have been confirmed by the application developers through our ethical reports.  
All the misuses are security-relevant and can lead to system crashes and Denial-of-service (DoS). \change{We plan to open-source our code and data later for future research\footnote{\url{https://github.com/icy17/GPTAid/}}.}
% As far as we know, we are the first to propose an approach for generating API APSRs using LLM.

\paragraph{Contributions.} We summarize the contributions as follows:

\vspace {3pt}\noindent$\bullet$\space\textbf{Novel technique.}
% 我们是第一个使用LLM生成APSR的工作。
We proposed a new approach to automate the APSRs generation using LLM.
% 我们的方法address了一些使用LLM需要解决的challenge包括：拆分复杂任务以重复利用LLM的推理能力， 使用执行反馈的方法来验证LLM生成结果的正确性，通过代码差异性分析的方法生成unambiguous的结果。
Our approach addresses two key challenges in directly using LLMs: the incorrect and overly-general APSRs arising. 
To solve these challenges, we adopt an execution feedback-checking approach to verify the correctness of the generated APSRs (Section~\ref{sec:method-validation}), and a code differential analysis is applied to generate the concrete APSRs with LLM (Section~\ref{sec:method-refine}). 
The generated APSRs are then proven to be effective through the application of API misuse detection, which outperforms the state-of-the-art detectors.

\vspace {3pt}\noindent$\bullet$\space\textbf{Insightful findings.}
We implemented \toolname{} on a subset of APIs from eight popular libraries. 
Among the generated 579 APSRs, we found 61.3\% of them have no corresponding description in the documentation, which means our work helps enrich the documentation. \change{We reported these APSRs to the library developers, and \confirmedAPSR{} of them have been confirmed.}
In total, all of the generated APSRs help detect \change{210} unknown security bugs, which could lead to severe security issues (such as system crashes\delete{ and Denial-of-Services (DoS)}), and \change{150} of them have been confirmed by developers after our ethical reports.
% 因此我们向API的开发者和使用者提出了一些建议来减少这种API misuse的发生。另外，在使用LLM是完成任务的过程中我们发现了一些prompt设计可能会显著影响LLM的效果，我们在XX提出了设计prompt的建议

\vspace {3pt}\noindent$\bullet$\space\textbf{Suggestions.}
Through the analysis of the \toolname{}'s performance, we got the chance to provide suggestions for API developers to enhance documentation for preventing API misuse (Section~\ref{sec:discuss-misuse}).
Besides, suggestions on prompt design to simplify the process and minimize errors generated by LLM are also provided in Section~\ref{sec:discuss-llm}.
\section{Background}
\label{sec:background}
% \begin{figure}
%     \centering
%     \includegraphics[width=0.95\linewidth]{Figure/prompt.pdf}
%     \caption{Prompt example}
%     % \vspace{-10pt}
%     \label{fig:bg-prompt-example}
% \end{figure}
% \subsection{\change{Application Programming Interface}}
% \todo{
% % API是
% }

\subsection{API Parameter Security Rules}
% API是一种软件库提供的可以被其他第三方软件直接调用的函数。通过使用API，其他软件的开发者不需要重复实现具有相同功能的函数，减少了开发过程中的负担。API在软件开发的过程中被广泛使用。通常软件库会向开发者提供API列表以及API文档，以说明API的具体功能以及使用API时的注意事项。In our research, 我们使用LLM生成第三方库的API。
\change{APIs are functions provided by libraries that other software can call directly, reducing repeated implementations and easing development~\cite{api-define}. They are widely used in software development, and libraries typically offer API lists and documentation. In our research, we use LLM to generate rules for third-party library APIs.}
% Parameter-related Security Rules (PRSRs) 描述了对API参数的使用限制。这个类型的规则包含参数值的限制以及对参数操作的限制。（1）参数值：API的参数作为输入，被用于API内部的不同的操作，developer必须保证向API传递的参数值是符合要求的（比如）。（2）对参数的操作：API的参数可能作为API的输入或者输出，当API作为输入时，developers需要保证该参数的状态满足某些条件，因此必须保证在调用该API前对其进行特定操作或者不进行特定操作（比如）。当API的参数作为API输出时，API内部操作改变了该参数的状态，developers在后续使用该参数时需要遵守操作限制（比如）。
API Parameter Security Rules (APSRs) define constraints on API parameters, addressing constraints on both parameter values and operations. \textit{(1) Parameter values}: 
API parameters serve as inputs for various operations within API, demanding that developers ensure compliance with specified values, such as the parameter must not be negative, \change{the value of parameter 1 must not be larger than the size of parameter 2, and the member of the parameter must not be NULL.} \textit{(2) Operations on Parameters}: 
\change{APIs can be used in complex call contexts, and the status of an API's parameters might be affected by multiple APIs. We define these types of rules as constraints on operations, such as the parameter must be freed later, the parameter must not be freed before, and so on.}
Violation of APSRs can result in security issues, such as crashes, memory corruption 
and denial of service. Therefore, detecting parameter-related API misuse is crucial for security. 
However, many APIs lack proper documentation, and creating APSRs manually is often hindered by limited expert knowledge and is time-consuming. Consequently, detecting parameter-related API misuses becomes a challenging task.
In our research, we use LLM to generate \change{various types of APSRs, 
such as value-related constraints, constraints among parameters, and constraints on operations. By applying these APSRs, \toolname{} can detect bugs caused by incorrect parameter use. For example, \toolname{} can detect null pointer reference (caused by incorrect parameter values), buffer overflow (caused by incorrect relationships among parameters), and memory leak (caused by incorrect operations on parameters).}
\subsection{Large Language Model}

% 什么是LLM（参数多，大量训练数据，基于XX方法，模型训练的来
% LLM是一类 在海量数据上训练的，有着非常庞大参数的artificial neural networks模型， 比如openai的gpt3模型参数量高达175 billion，训练数据集来自维基百科，书以及网络上的数据。大量的训练数据及参数使LLM可以不经fine-tune地应用在特别的领域上
Large Language Models (LLMs), like OpenAI's GPT-3 with 175 billion parameters~\cite{gpt}, are neural networks trained on vast datasets. This extensive training allows LLMs to perform specific tasks without fine-tuning. 
% LLM like GPT通过prompt与用户进行交互。prompt是用户提供给LLM的输入，其中包含着用户对于需要完成任务的描述，比如图XX中所示，prompt包含着任务指令：计算等式结果，LLM会理解prompt中描述的任务并给出答案。
LLMs like GPT interact with users through prompts, which are user-provided inputs describing the task to be accomplished\delete{, for example, as shown in Figure~\ref{fig:bg-prompt-example}, the prompt contains the task instruction: generate parameter-related security rules for \texttt{free}, and the LLM understands the task described in the prompt and gives an answer}.
% 为了提升LLM在不同领域问题上的推理能力，现有研究提出了一些设计prompt的方法。zero-shot是一种不在prompt种提供交互示例的prompt设计思想。few-shot通过在prompt中提供交互示例对LLM进行引导。Chain-Of-Thoughts是state of the art的通过设计step-by-step的步骤提升LLM推理能力的方法。基于Chain-Of-Thoughts思想，将需要完成的任务拆分成多个中间步骤，并引导LLM按步骤完成任务可以提升LLM的推理能力。现有方法通过结合zero-shot/few-shot与CoT获得更好的结果。
To enhance the reasoning ability of LLMs on various specific tasks, existing research has proposed some approaches for designing prompts (known as \textit{prompt engineering}).
Zero-shot prompts~\cite{wei2022zeroshot} do not include interaction examples. Few-shot prompts~\cite{brown2020fewshot} guide LLMs by incorporating examples. The Chain-of-Thought~\cite{wei2023cot} approach represents the state-of-the-art technique, enhancing reasoning through step-by-step design.
% LLM（gpt3.5）训练集XXX 因此LLM具有XXX能力
% LLM已有应用
Previous approaches~\cite{dengllmfuzz2023,  xiallmfix2023, aprXia2022} used LLMs for software security. In our research, LLM is employed to generate APSRs. 
% 然而由于幻觉问题，LLM生成的结果可能存在错误，如何识别错误的结果是非常困难的。
However, the outputs of LLMs can be incorrect or overly general, 
\jht{which can not be easily resolved by mere prompt engineering.
Identifying these incorrect or general responses of LLMs is quite challenging.}

\section{Methodology} 
In this section, we introduce \toolname{}, which is designed for automatically generating accurate and concrete API Parameter Security Rules (APSRs) using LLM and detecting parameter-related API misuses. We start with the overview and use an example to illustrate the workflow of \toolname{}, followed by a detailed description of each component. 
\subsection{Overview}

\noindent\textbf{Architecture. }
\begin{figure}
    \centering \vspace{-6pt}\includegraphics[width=0.5\textwidth]{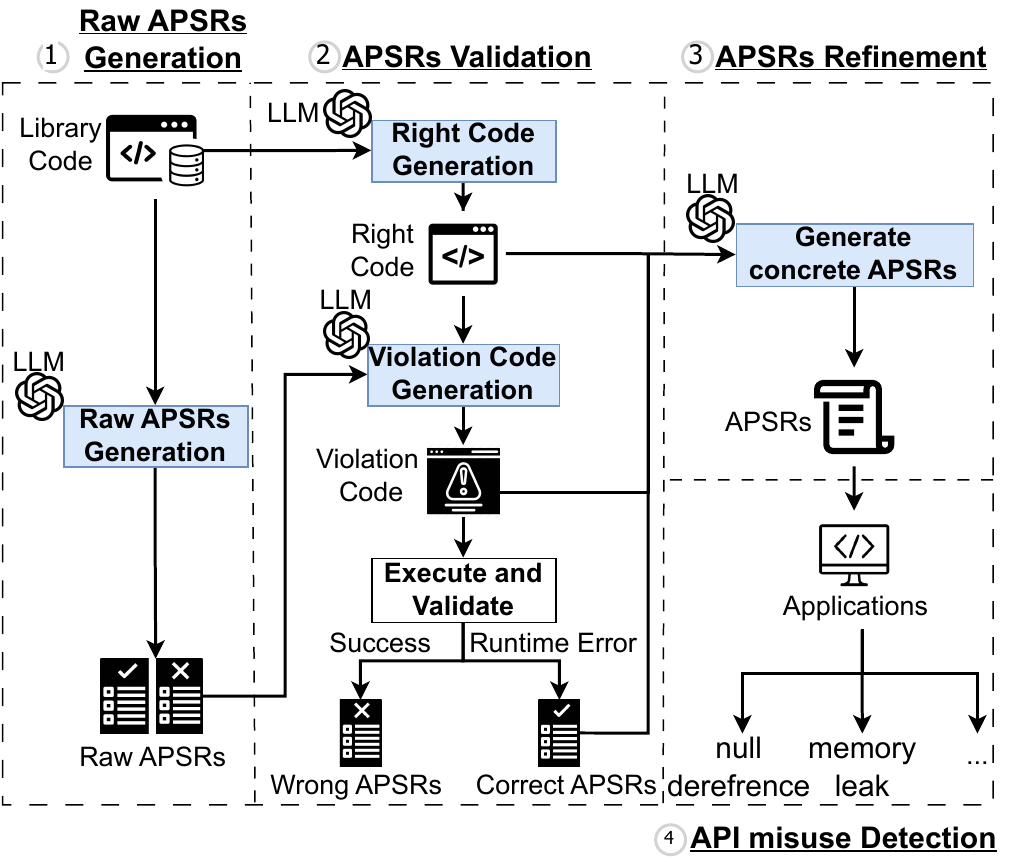}
    \caption{Architecture of \toolname{}}
\label{fig:arch}
\vspace{-6pt}
\end{figure}
% figure 包含两个部分, XXXX。XX包含XXX, 。。。
Figure~\ref{fig:arch} illustrates the architecture of \toolname{}, consisting of four stages: Raw APSRs Generation, APSRs Validation, APSRs Refinement and API misuse detection. 
In Raw APSRs Generation (stage-1), \toolname{} automatically constructs a prompt with the API source code and prompts LLM. This enables LLM to analyze the API source code and generate raw APSRs. However, the generated raw APSRs might be incorrect. Therefore, for each raw APSR, \toolname{} validates the correctness of it based on execution feedback (stage-2).
In this stage, \toolname{} first instructs LLM to generate the right code ($C_r$) calling the target API. To ensure $C_r$ executes without triggering any runtime error, \toolname{} monitors execution feedback and automatically repairs runtime errors. Subsequently, for each raw APSR, \toolname{} automatically instructs LLM to modify the $C_r$ to generate the violation code ($C_v$) that violates the target raw APSR. \toolname{} then automatically executes the $C_v$ and monitors if a runtime error occurs to identify the correct raw APSRs.
To enable the use of APSRs for API misuse detection, \toolname{} refines all the correct APSRs to generate concrete APSRs (stage-3). For this purpose, \toolname{} instructs LLM to analyze the difference of $C_r$ and $C_v$ to generate APSRs related to the modification operations. 
The APSRs generated are used to detect API misuse with CodeQL~\cite{codeql} (stage-4).

\noindent\textbf{Example. }
\label{sec:methodlogoy}
\begin{figure}
    \centering
    \vspace{-6pt}\includegraphics[width=0.9\linewidth]
    {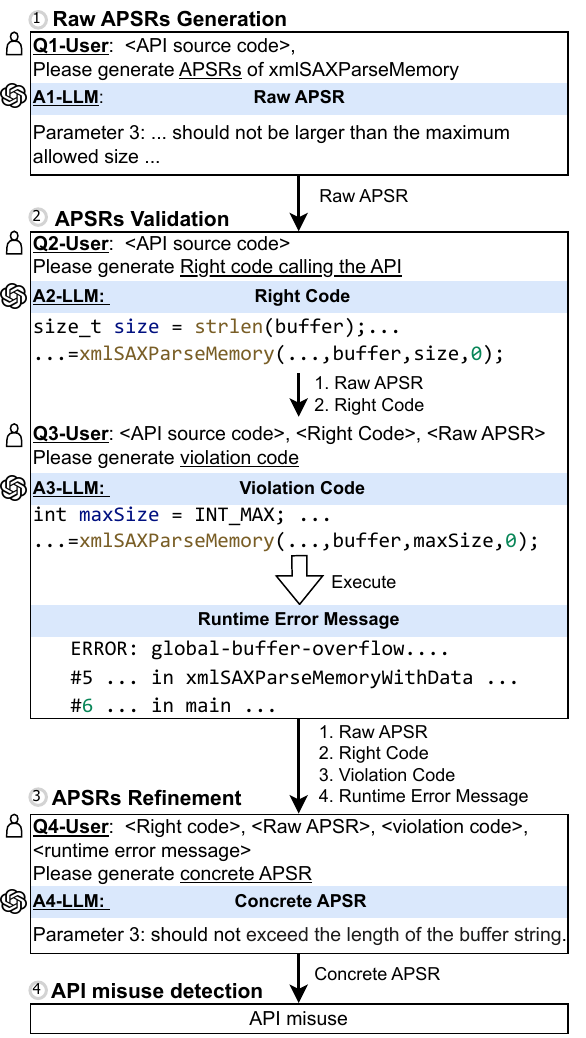}
    \caption{An example of \toolname{}'s workflow} 
\label{fig:arch-example}
\vspace{-5pt}
\end{figure}
% sqlite3_open:right:valid, wrong: non exists db  3个过程
The example in Figure~\ref{fig:arch-example} introduces the workflow of \toolname{}. 
It describes the entire process of \toolname{} generating the APSRs of API \texttt{xmlSAXParseMemory}, an API in the libxml2 library that ``\textit{parse an XML in-memory block and use the given SAX function block to handle the parsing callback}''~\cite{libxml2doc}.
First, \toolname{} provides a prompt containing source code of \texttt{xmlSAXParseMemory} to LLM and LLM generates a raw APSR: \textit{``Parameter 3: The size parameter should not be larger than the maximum allowed size to prevent denial-of-service attacks or memory exhaustion.''} as is shown in \textit{A1}. Then \toolname{} instructs LLM to generate the $C_r$ and then modify it to generate $C_v$ that violates the raw APSR. To generate the violation code, LLM modifies the third parameter from \texttt{strlen(buffer)} to \texttt{INT\_MAX}, shown as the $C_r$ and $C_v$ in \textit{A2} and \textit{A3}, respectively. Subsequently, \toolname{} executes the $C_v$ and catches a runtime error using the sanitizer~\cite{asan}, confirming the correctness of this APSR. \toolname{} then instructs LLM to analyze the key modification operations by identifying the difference between the $C_r$ and $C_v$ and generate a concrete APSR that describes a relation between the values of parameter 2 and parameter 3, as shown in \textit{A4}. Finally, the concrete APSR is used to detect API misuse.

\subsection{Raw APSRs Generation}
\label{sec:method-generation}
\begin{figure}
\vspace{-6pt}
    \centering
    \includegraphics[width=1\linewidth]{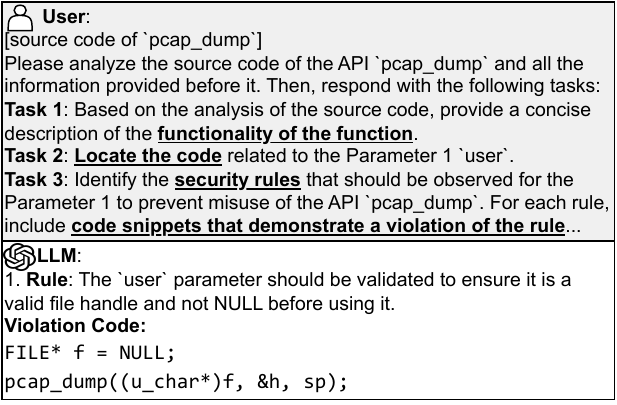}
    \caption{Prompt example of Raw APSRs Generation} 
\label{fig:prompt-rule}
\vspace{-5pt}
\end{figure}
This stage aims to generate APSRs by analyzing API source code. 
Most of the existing work~\cite{Lyugoshawk2022, huaurc2023, ZhouAnalysing2017, Zhongempir2020} is limited to detecting specific types of API misuse, relying on specific code analysis rules for source code analysis. \jh{Manually constructing these code analysis rules is very time-consuming and limited by expert knowledge.
LLMs are powerful tools that have the ability to comprehend code without the need of code analysis rules.}
Therefore, \toolname{} uses LLM to generate APSRs by analyzing API source code automatically. The APSRs generated in this stage are referred to as \textit{raw APSRs}.

Initially, instructing LLM to generate APSRs by directly analyzing the source code of APIs may seem straightforward. However, this simple approach dose not work well  
when dealing with intricate API source code that involves multiple parameters and complex implementation logic. 
Analyzing multiple targets from a large amount of code is challenging for LLM, causing missing rules. 
\jh{For example, when we provide LLM with the source code of \texttt{pcap\_dump} -- an API with three parameters -- and instruct it to generate all related APSRs, it generates only one correct APSR, missing four others.}
To address this issue, we propose a method that break down the task.
% 为了拆分任务，我们首先通过静态分析的方式确定目标API的参数信息，然后将生成API参数相关的安全规则这一任务分解为分别生成单独参数的安全规则，通过这种任务拆分，可以让LLM每次只针对一个目标参数进行分析。对于每个参数，我们基于chain of thought的思想，将问题拆分成多个中间步骤以提升LLM的推理能力。我们使用step by step的prompt来引导llm生成该参数相关的安全规则。
To break down the task, \toolname{} begins by identifying the numbers and names of parameters of target API through static analysis. \toolname{} then instructs LLM to generate raw APSRs for each parameter rather than generating raw APSRs for all the parameters. This enables LLM to focus on one parameter at a time and simplifies the process of linking APSRs with their related parameters.

% {Figure/arch_example-v2.pdf}
\noindent\textbf{Prompt Design. }
% 我们对比了三种不同的prompt engineering方法，我们发现few-shot往往会受限于example中生成的规则类型，难以生成多样化的规则,因此我们选择了效果相对更好的chain-of-thought的方式设计prompt.
To fully utilize LLM for generating raw APSRs, we evaluated three prompt engineering methods: zero-shot, few-shot, and Chain-Of-Thought. Our analysis showed that Chain-Of-Thought performed best, so we adopted it for prompt design. Detailed comparisons of them are provided in Appendix~\ref{sec:temp-selection}. 
The prompt consists of two parts: the API source code as input information and step-by-step instructions for the LLM \change{to complete}, as shown in Figure~\ref{fig:prompt-rule}. 
\change{
We design each step's instruction based on the principle that the LLM can complete the step by analyzing the information given in the prompt, and that completing the step will aid the LLM in generating the APSRs.
}
The instruction is divided into the following three steps:
\ignore{summarize the functionalities of the API, locate the parameter-related lines of code and generate raw APSRs with the violation code example that violates the raw APSRs.}
\\
\ding{182} \textbf{Summarizing API functionalities} helps LLM to mine potential APSRs.
There is a relationship between API functionalities and its potential APSRs. For example, when an API has memory allocation functionality, the corresponding APSR is the need to release allocated memory. 
\\
\ding{183} \textbf{Locating parameter-related lines of API source code} reduces the amount of code that LLM needs to analyze. This step helps the LLM reduce the impact of irrelevant code on the correctness of generating the APSRs.
\\
\ding{184} \textbf{Generating raw APSRs with their violation code examples} can express the raw APSRs in a straightforward way. As mentioned earlier, the raw APSRs generated by LLM are overly general causing the gap between raw APSRs and downstream tasks.
% for example，图中的APSR是“”，人在分析该规则时可能产生不同的理解如图XXX所示。通过分析源码我们发现，这里APSR代表的真正含义是不能为NULL，然而由于general，可能将其理解为错误的内容。
For example, as previously mentioned, the general APSR is: \textit{``Parameter 2: The \texttt{ppDb} parameter must be a valid pointer to an \texttt{sqlite3} structure before calling \texttt{sqlite3\_open}''}. \jh{There are three possible interpretations of \textit{valid} in this APSR, but only one is correct. }Overly general APSRs can result in errors during subsequent use. 
To solve this problem, it is essential to identify information that can represent a specific constraint of the API and use it as the supplementary information to the raw APSRs.
% 因为编程语言是一种形式化语言, 相对于自然语言来说不容易产生歧义, 因此我们选择使用代码来作为规则的补充信息。为了使规则之间的区分度足够明显, 我们需要选择一个表示合适信息的代码作为补充信息, 能够尽可能区分不同的规则。
Programming language is a formal language that is more concrete compared to natural language.
Intuitively, for different APSRs, the code violating these APSRs is distinct and the violation pattern may provide insights into the constraints of the API. Based on this assumption, a violation code example can be used to represent a specific constraint of API.
\jh{Therefore, in the last step of the instruction, we instruct LLM to generate raw APSRs with their violation code examples, providing supplementary information for these raw APSRs. }In our evaluation, \toolname{} achieved a recall of \ablationRawRecall{} in generating raw APSRs, generating more APSRs than other approaches (Section~\ref{sec:compare-study}).

% Table generated by Excel2LaTeX from sheet 'Sheet1'

% %%%%%%%%%%%%%%%%%Validation%%%%%%%%%%%%%%%%%%%%%%%%%%%%%%%
\subsection{APSRs Validation}
\label{sec:method-validation}
\begin{figure}
\vspace{-6pt}
    \centering
    \includegraphics[width=0.8\linewidth]{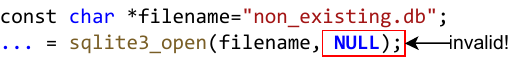}
    % \normalsize
    \caption{False violation code leading to unrelated API misuse}
    \label{fig:right-example}
    % \vspace{-8pt}
\end{figure}
% 由于LLM存在Hallucination问题，LLM可能给出错误的答案。因此需要对LLM的结果进行正确性检验。
As mentioned earlier, LLM may produce incorrect answers. 
This stage aims to validate the correctness of the raw APSRs generated by LLM in the previous stage, which contains challenges.  
% 由于不存在绝对正确的参考信息
The raw APSRs are expressed in natural language, and automating the assessment of their correctness can be quite challenging due to the absence of definitive reference information. 
% 通过分析已有API误用，我们发现94%的API误用都会导致执行时出现runtime error或者被sanitizer检测到，因此
After analyzing the reported API misuse in prior studies~\cite{jiang2024appminer, Lyugoshawk2022, lin2023aphp, lv2020rtfm}, we identified that 94\% of them lead to runtime errors during execution which can be caught by monitoring tools (Table~\ref{tab:errmsg-misuse} in Appendix). Therefore, we assume that if a raw APSR is correct, the code that violates the raw APSR will cause runtime errors. Based on this assumption, \toolname{} validates the correctness of a raw APSR by analyzing the execution result of the code violates \dels{the}it. 
In a few cases, violating an APSR does not lead to runtime errors, causing \toolname{} to fail in generating the corresponding APSR. We will discuss this limitation in Section~\ref{sec:discuss-limitation}.

Directly generating API calling code that violates the raw APSRs might lead to API misuse unrelated to the raw APSRs. 
% 不能区分API misuse是否是由于违反该raw APSR导致的会导致基于执行结果的分析错误地将错误的raw APSR验证为正确的。
Failing to differentiate whether API misuse results from a violation of the raw APSR can lead to the execution-based analysis incorrectly confirming an incorrect raw APSR as correct.
For example, if LLM is instructed to directly generate code that violates the raw APSR of \texttt{sqlite3\_open}: \textit{``Parameter 1: The filename should be validated to ensure it refers to a legitimate, existing database file''}, LLM might generate code that looks like the code in Figure~\ref{fig:right-example}. While this code does violate the APSR successfully, it passes the second parameter as NULL which leads to a runtime error unrelated to this raw APSR. Based on the assumption, this raw APSR is categorized as correct, which leads to an incorrect APSR. It is challenging to distinguish these incorrect violations due to the absence of the correct API usage patterns.
To solve this problem, \toolname{} first generates the right code ($C_r$) calling the API without runtime errors and modifies the $C_r$ according to raw APSRs to get the violation code ($C_v$) that violates the raw APSRs. Through the modification, \toolname{} ensures the runtime errors of $C_v$ are related to this modification based on raw APSRs. Finally, \toolname{} verifies the correctness of raw APSRs by executing the $C_v$ and analyzing if there are runtime error messages (\textit{REMs}) related to the target API. 
This stage consists of three steps: right code generation, violation code generation, and correct APSRs verification. We delve into the details of these steps below. 

\noindent\textbf{Right Code Generation. }
\label{sec:right-code-generation}
\begin{figure}
    % \vspace{-6pt}
    \centering
    \includegraphics[width=0.88\linewidth]{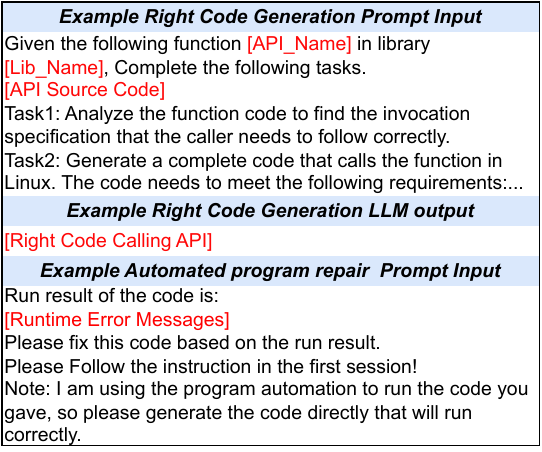}
    \caption{Prompt example of Right Code Generation}
    \label{fig:prompt-right}  
\end{figure}
The aim of this step is to generate $C_r$. 
% % 考虑到复杂代码可能难以编译运行成功，并且过多的无关代码会增加分析结果的难度。
% Complex code can make compiling and running it successfully a challenge, while too much irrelevant code can complicate result analysis.
% The $C_r$ should be as simple as possible to minimize the introduction of API-unrelated factors.
Existing approaches primarily mine the API usage patterns from extensive code corpora and subsequently generate API-calling code~\cite{fuzzgen2020, fudge2019}. However, these approaches have a low success rate in generating code, and they often produce redundant code. 
Complex code can make compiling and running it successfully a challenge, while too much irrelevant code can complicate result analysis.
The $C_r$ should be as simple as possible to minimize the introduction of API-unrelated factors.
Because LLMs' capacity to generate API calling code~\cite{dengllmfuzz2023}, supported by its vast knowledge of code, \toolname{} utilizes LLM for generating $C_r$ that calls API. 

% 指导LLM生成正确代码的prompt由两部分组成：API源码和step by step的instruction, 这个instruction包含两个任务：根据源码分析调用该API要注意的使用规范以及生成正确调用代码。
To generate $C_r$, we design a prompt consisting of two parts: the API source code and step-by-step instructions.
The step-by-step instruction contains two tasks: analyzing specifications for calling the API correctly based on the API source code and generating the $C_r$.
% 由于正确调用某些API要求构建较为复杂的上下文环境, 保证LLM能够正确生成满足调用上下文的代码是非常困难的.
\change{Since correctly calling certain APIs requires constructing complex contexts, it is challenging for LLM to generate API calling code that satisfies these contexts accurately.} To address this, \toolname{} instructs LLM to perform automated program repair when compile errors or runtime errors arise.
\change{In this way, \toolname{} can generate the correct calling code by modifying the incorrect context and gradually satisfying the complex context step-by-step.}
Specifically, \toolname{} provides LLM with the session history that is used to generate the $C_r$, with any error feedback from execution and instructs LLM to fix code errors. 
% 一旦执行目标API成功，或者自动化修复到达最大的conversation 限制，自动化修复停止。
\jh{When the execution of $C_r$ is successful or the automated repair reaches the maximum repair times, the automated repair process stops. The prompt example of Right Code Generation is shown in Figure~\ref{fig:prompt-right}. }Our study shows the effectiveness of \toolname{}, achieving a success rate of \rightSuccessRate{} in generating the correct API calling code.

\begin{figure}
    \centering
    \vspace{-6pt}\includegraphics[width=0.75\linewidth]{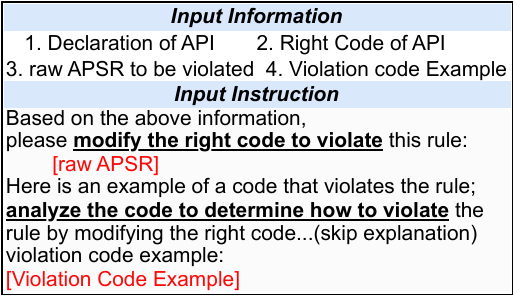}
    \caption{Prompt template of Violation Code Generation} 
\label{fig:prompt-violation}
\vspace{-6pt}
\end{figure}

\noindent\textbf{Violation Code Generation. }
The aim of this step is to generate the violation code ($C_v$) that violates raw APSRs by modifying the $C_r$ based on the provided raw APSRs. 
To achieve this, \toolname{} instructs LLM with a prompt containing a task description and four parts of information including the $C_r$, API declaration, raw APSR, and violation code example for raw APSR. The API declaration aids the LLM in analyzing the APSR and identifying the target parameter described within. The violation code example helps to provide specific details and clarify the APSR.
\change{By following the instructions, \toolname{} can modify $C_r$ to $C_v$ in various ways, such as changing parameter values or altering the API calls in the calling context involving the same variable.}
\jh{The prompt template is shown in Figure~\ref{fig:prompt-violation}.}
Similar to right code generation, $C_v$ can contain errors and cannot run. We use the same repair method to solve this problem. 
As previously mentioned, the output generated by LLM may contain errors. In this step, the error could be that LLM may not correctly modify the $C_r$ according to the instructions leading to the APSR-unrelated modification. 

% 这种错误在这一步表现在不按照APSR的要求进行修改。
% 为了解决这个问题, 我们通过分析代码修改的位置与规则描述的内容是否一致来检验修改是否存在问题。
To address this, \toolname{} roughly checks whether the modifications are wrong by automatically analyzing whether the modifications of the code are consistent with what the raw APSRs describe. 
% 我们首先通过静态代码分析技术分析right code和修改后代码的区别, 确定修改的参数Cpara及修改处与目标API之间的位置关系Cloc. 然后通过分析raw APSR的关键词, 从raw APSR中提取该raw APSR描述的参数Rpara以及描述的动作相对目标API的位置关系Rloc。最后通过对比（Cpara, Cloc）与（Rpara, Rloc）是否一致来实现对修改结果正确性的检验。
\toolname{} first automatically identifies the differences between the $C_r$ and the modified code. \toolname{} then utilizes abstract syntax tree (AST) analysis to pinpoint target API calls and which parameter is associated with the modified code snippets. This process identifies the modified parameter ($C_{para}$) and the location relation ($C_{loc}$) between the modified code snippets and the target API. Meanwhile, \toolname{} analyzes the raw APSRs using location keywords (e.g., before) to extract the location relation ($R_{loc}$) of the described action relative to the target API and gets the target parameter ($R_{para}$) directly from the Raw APSRs Generation stage. \toolname{} assesses the correctness of the modification by comparing the consistency of ($C_{para}$, $C_{loc}$) and ($R_{para}$, $R_{loc}$). 
% 与right code generation类似, 修改后的violation code可能存在无法运行的问题。我们使用了相同的APR方法解决该问题。
% \todo{Example?}
% 不确定是否需要给个静态分析的例子, 先不加

\begin{figure}
    \centering
\vspace{-6pt}    \includegraphics[width=0.9\linewidth]{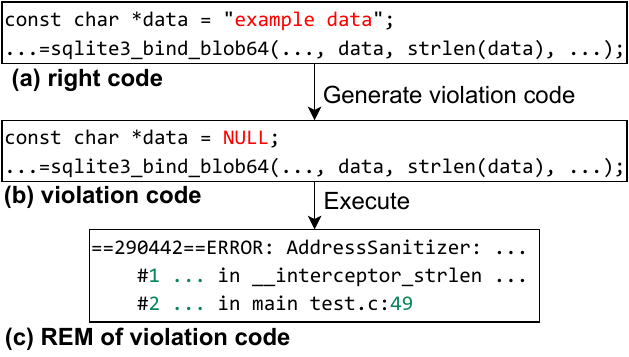}
    \caption{Violation code leading to an unrelated bug}
    \label{fig:validation-example}
\end{figure}

\noindent\textbf{Correct APSRs verification. }
This step aims to verify the correctness of the APSRs based on the runtime execution output of $C_v$. As previously mentioned, the assumption is that an APSR is considered correct if the $C_v$ leads to runtime errors during execution. 
However, runtime errors might be unrelated to the target API, 
\ignore{Unfortunately, we can not simply presume that the rule corresponding to the code where the runtime error occurs is inherently correct. We need to verify that the error is indeed caused by a violation of that rule. Even if we conducted static analysis on the violating code in previous steps, there is still a possibility that the violating rule might affect other code simultaneously, leading to runtime errors in the unrelated code. }
as shown in Figure~\ref{fig:validation-example}, the $C_r$ to call the API \texttt{sqlite3\_bind\_blob64} is in Figure~\ref{fig:validation-example}(a), and the APSR to be checked is: \textit{``Parameter 3 must not be NULL''}. To violate this APSR, LLM modifies the $C_r$ to the $C_v$ in Figure~\ref{fig:validation-example}(b). This modification sets the variable \texttt{data} to NULL, precisely violating the APSR. As Figure~\ref{fig:validation-example}(c) shows, the violation code causes a runtime error. 
However, the variable \texttt{data} is also passed to \texttt{strlen}. Unfortunately, passing a NULL parameter to the \texttt{strlen} function leads to a null pointer dereference, resulting in a runtime error unrelated to the API. 

To distinguish between errors caused by the target API and those caused by unrelated factors, we design a process to analyze REMs \change{generated by the monitors} automatically. 
\change{
REMs describe where an error occurred in the code and provide a stack trace showing the call sequence leading to the error. First, \toolname{} analyzes the error trace to determine if the error occurs at the location of the code where the target API is called. Then, to further determine if the error is API-related, \toolname{} examines the subsequent call sequence in the error trace to see if the error occurs in the implementation of the target API.}
\delete{First, \toolname{} analyzes the stack frames in \texttt{main} function in REM, to check whether the error originated within the line of code calls target API. Then, \toolname{} analyzes the stack trace frames above the \texttt{main} function in REM to check whether the code in the trace is in the target API.} For example, in the Figure~\ref{fig:validation-example}(c),  \toolname{} first identifies the error in the \texttt{main} function (\texttt{main test.c:49}), where there is a line that includes function calls to \texttt{sqlite3\_bind\_blob64} and \texttt{strlen}. This indicates that the error is very likely caused by the API. \toolname{} then analyzes the stack trace frames above \texttt{main} to confirm whether the error originated within the API rather than from \texttt{strlen}. The frame above is: \texttt{in \_\_interceptor\_\_strlen}, indicating that this REM is related to \texttt{strlen}, not the API. Based on the above analysis, \toolname{} can ascertain whether the REM is related to the API. 
%%%%%%%%%%%%%%%%%%%%%%%%%%%%%Refinement%%%%%%%%%%%%%%%%%%%%%%%%%%%%%%%%%%%%%%%%%%%%%%%%%%%%%%%%
\subsection{APSRs Refinement}
\label{sec:method-refine}
\begin{figure}
    \centering
    \includegraphics[width=0.95\linewidth]{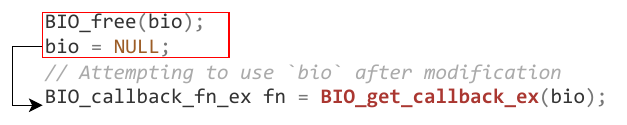}
    \caption{Modification contains multiple operations}
    \label{fig:refine-example}
    \vspace{-5pt}
\end{figure}

% 在过滤掉了错误的APSR后，XX is ready to让APSR更加concrete以能够被用来自动化地生成准确的检测代码。
As previously mentioned, the APSRs generated by LLM are general, making them unsuitable for use as reliable information for API misuse detection. Therefore, refining APSRs to express concrete constraints becomes crucial. 
% 基于这样的observation：code更加直接不容易导致歧义，我们考虑通过分析代码来生成与代码行为一致的APSR。
Considering code is more straightforward and concrete (Section~\ref{sec:method-generation}), we propose an approach based on code differential analysis to guide LLM in generating APSRs consistent with code. 
% 考虑到violation code是在right code的基础上通过修改得来的, 而这种修改导致了API misuse, 因此安全规则应该清晰明确地要求caller避免出现修改中使用的API调用方式。正如SectionXX提到的, 代码作为一种形式化语言表达的含义是更为清晰的。因此我们通过分析正确代码与错误代码之间的差异来生成一义化的APSR。
Specifically, the $C_v$ is a modification of the $C_r$ and the modification leads to API misuse. Well-defined APSRs should contain constraints aimed at preventing incorrect API usage introduced by these modifications. Based on this phenomenon, analyzing the modifications between the $C_r$ and the $C_v$ and generating APSRs that describe these modifications can be helpful in producing concrete APSRs. 
However, when analyzing code modifications involving multiple operations, LLM may struggle to identify the specific modification operations directly related to API misuse. This issue could lead to LLM generating an incorrect APSR based on the unrelated operations. 
For example, as shown in Figure~\ref{fig:refine-example}, for API \texttt{BIO\_get\_callback\_ex}, the generated $C_v$ is shown in the figure, which results in an error. 
% 图中展示了violation code与right code不同的地方在于调用目标API前free了该变量并将其赋值为NULL。
The difference between $C_v$ and $C_r$ is that $C_v$ frees the \texttt{bio} parameter and then sets it to NULL before calling the target API.
The error is caused by the passing NULL as the parameter, and the error is not related to the \texttt{free} function. In this case, identifying the exact operation is challenging due to the multiple modifications involving both the \texttt{free} function and the NULL pointer. 

To solve this problem, \toolname{} first groups different violation code that lead to the same runtime error, and then identifies the shared operations as the key operations. Specifically, we observe that the same API-related runtime errors typically result from the same operations in the code.
Therefore, \toolname{} analyzing different $C_v$ that leads to the same API-related runtime errors, aiming to identify common code modification operations shared among them.
This helps \toolname{} pinpoint the key operations causing the API misuse from several modification operations.
For this purpose, \toolname{} begins by grouping $C_v$ that share the same REMs. 
To efficiently group REMs, \toolname{} focuses on essential information within the REMs while discarding irrelevant factors, such as process IDs, which could distort clustering. Specifically, it identifies the potential causes described in the REMs and analyzes the associated error stack trace \change{to identify the call sequence using the same method as in APSRs validation}. Then, it groups REMs based on these crucial details.
Subsequently, \toolname{} provides LLM with information from each cluster, instructing LLM to analyze differences between all the $C_v$ and their $C_r$ to identify the modification operations in code. \toolname{} then instructs LLM to analyze the common modifications operations among them to identify the key operations. Finally, \toolname{} instructs LLM 
 to generate the APSRs containing descriptions of the key operations for each cluster. 

\noindent\textbf{Prompt Design. }
\begin{figure}
    \centering
\vspace{-6pt}
\includegraphics[width=1\linewidth]{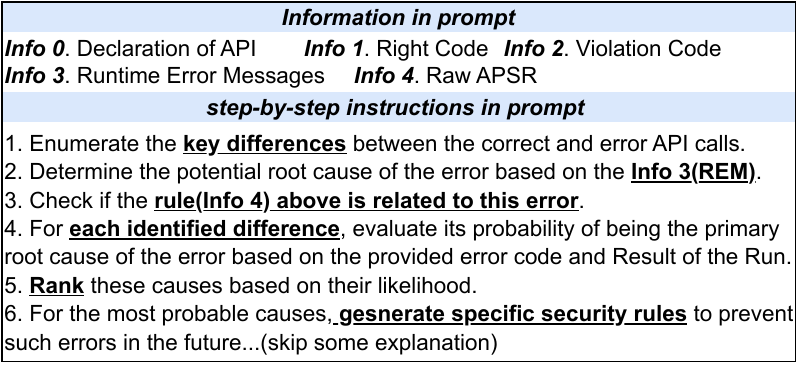}
    \caption{Prompt template of APSRs Refinement} 
\label{fig:prompt-refinement}
\vspace{-6pt}
\end{figure}
% implementation?
To enable LLM to analyze all available information and generate concrete APSRs based on the key operation, we design a prompt \change{consisting} of information and step-by-step instructions \change{that LLM needs to complete}. There are six main steps of instructions: identify differences, analyze shared REMs, analyze raw APSRs, analyze differences, rank the possibilities and generate APSRs. The prompt template when only one $C_v$ in a cluster is shown in Figure~\ref{fig:prompt-refinement}. 
\ignore{The prompt template when multiple $C_v$ in a cluster is shown in Figure~\ref{fig:cprompt-refine-multi} in appendix.}
Specifically, \ding{182} Identifying differences between each $C_r$ and $C_v$ pair in a group helps the LLM pinpoint all modification operations linked to the same runtime error. \ding{183} Analyzing shared REMs enables the LLM to identify potential causes of API misuse as described in the REMs. \ding{184} Analyzing raw APSRs enables the LLM to identify potential API misuse causes as described in the APSRs. This step is omitted for the cluster with multiple codes, as their distinct raw APSRs are less likely to directly relate to the root cause. 
\ding{185} Analyzing all the differences and identify one root cause helps LLM to identify key operations lead to the API misuse among numerous modification operations. 
\ding{186} Ranking the possibilities allows the LLM to evaluate all potential causes identified previously and determine the most likely cause of API misuse. \ding{187} In the generating APSRs step, \toolname{} instructs LLM to generate concrete APSRs based on the operation and the most likely cause.
% 为了使LLM能够基于代码的不同生成APSR, 我们在prompt中包含了right code和violation code并设计任务引导LLM分析两段代码的不同。为了让LLM生成与root cause相关的APSR, 我们在prompt中包含了REM并引导LLM通过分析REM总结出可能存在的问题。
% To guide LLM in generating APSRs based on code differences, \toolname{} provide the $C_r$ and $C_v$ in the prompt. We then design a task instructing LLM to analyze the differences between these two code. 
% To ensure that LLM generates APSRs related to the API-related runtime errors, we include REMs in the prompt and instruct LLM to analyze REMs to summarize the potential issues. 
% % 当基于REM的类中只有一个violation code时, 我们将该violation code对应的raw APSR提供给LLM, 引导LLM分析违反该raw APSR可能导致的问题并作为root cause的可能性备选。
% When there is only one $C_v$ in a cluster, we supply the raw APSR that corresponds to the $C_v$ to LLM. We instruct LLM to analyze the potential issues arising from violating the raw APSR, considering it as a potential alternative to identifying the constraints. 
% % 当基于REM的类中有多个violation code时, 考虑到多个violation code对应的raw APSR是不同的, 这些raw APSR与root cause相关的可能性较低, 因此对于这种情况我们不会在prompt中提供raw APSR, 而是引导LLM通过分析多对代码对差异的相同点分析root cause。
% When multiple $C_v$ exist within a cluster, and given that the raw APSRs corresponding to these codes are distinct, they are less likely to be directly related to the root cause. Therefore, instead of including raw APSRs in the prompt, we guide LLM to analyze the constraints by analyzing common modification operations. 
% prompt的具体内容我们在Section XX详细描述
In this way, APSRs are refined and can be used to generate detection rules for API misuse detection. Our experiment shows \toolname{} achieves an accuracy of \ablationFinalPre{} in generating APSRs (Section~\ref{sec:ablation}).
\subsection{API misuse detection}

\begin{figure}
    \centering    
    \vspace{-6pt}
    \includegraphics[width=1\linewidth]{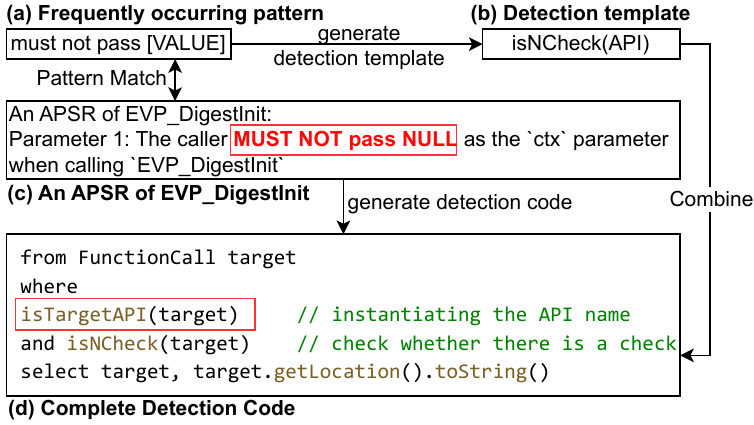}
    \caption{An example of detection code generation} 
\label{fig:ql-generation-example}
\vspace{-6pt}
\end{figure}
% 我们对所有的APSR进行了人工分析并识别出了频繁出现的pattern，我们将这些pattern分为三个大类。它们分别是Value, Relation, Action. Value指的是对parameter值的限制，比如API的APSR：parameter 1 must not be NULL. Relation 指的是对多个parameter关系之间的限制，比如APIXXX的APSR：Parameter 3: The value of the `size` parameter should not exceed the length of the provided buffer(parameter 2).Action指的是在调用API前或者调用API后必须进行的或必须不进行的动作，这个动作可能指的是initialize, free, use, malloc。比如APIxxx的APSR：parameter must not be freed before. 我们为Value和Relation类各设计了一个template，针对Action类，不同的动作，我们设计了不同的template。

In this stage, we use APSRs for API misuse detection. Our approach draw inspiration from Advance~\cite{lv2020rtfm}, which detects API misuse through security rules (referred to as IA in Advance) using CodeQL~\cite{codeql}. CodeQL is a tool for static analysis on target applications based on detection QL code composed using detection rules. 
% Advance通过frequent subtree mining将IA中的predicate聚类，并为不同的predicate人工构建检测代码template，最后通过组合这些predicate为每一条IA自动化生成检测代码。由于Advance并未开源聚类与自动化生成检测代码的源码，我们参考Advance的方法人工为APSR中频繁出现的模式生成template并基于template自动化生成APSR对应的检测代码。
Advance employs Natural Language Processing (NLP) techniques for clustering predicates in IA through frequent subtree mining, manually constructs detection code templates for each frequently occurring predicate, and finally automatically generates detection code for each IA by combining these templates. We start by manually identifying the frequently occurring description patterns across all APSRs and creating detection code templates for these patterns. Subsequently, \toolname{} automatically combines these templates and instantiates the parameter index and API name to generate the complete detection code (QL code) and employs this code for API misuse detection using CodeQL.
% For example, as shown in Figure XX. 通过人工分析APSR，我们得到了频繁出现的pattern之一是must not be [VALUE]，其中value可以被任意值替代。我们为这个pattern人工生成的template为isCheck函数，该函数使用控制流分析，检测在调用目标API前，是否对目标参数进行过检查。图XX可见 API xxx的APSR中有must not be NULL,匹配了该pattern，因此\toolname{}通过自动化结合该template并实例化API名和参数得到最后的检测代码如图XX，完整代码见附录XX。
\jh{
For example, the process of detection code generation is shown in Figure~\ref{fig:ql-generation-example}. Through our manual analysis of APSRs, we identify one of the frequently occurring description patterns is \textit{``the caller must not pass [VALUE]''} (Figure~\ref{fig:ql-generation-example}(a)), where \texttt{VALUE} can be any value. We manually created a detection code template for this pattern ({isNCheck} function in Figure~\ref{fig:ql-generation-example}(b)). This function employs data/control flow analysis to ascertain whether the target parameter has  been checked before the API is called. 
Figure~\ref{fig:ql-generation-example}(c) shows the APSR for API \texttt{EVP\_DigestInit}, specifying that \textit{``the caller must avoid passing NULL''}, which matches the identified pattern. \toolname{} automatically combines the detection code template (isNCheck) and instantiates the API name and parameter index to generate the  detection code, as depicted in  Figure~\ref{fig:ql-generation-example}(d). \delete{The complete code is available in Listing~\ref{lst:ql-example} in the Appendix.}
}

\section{Implementation}
\label{sec: implementation}
In this section, we detail the implementation of the components of \toolname{}.
% 我们使用了LLM的gpt-3.5-xxx来完成所有需要LLM的任务。我们将所有使用LLM的任务的temperature设为1来允许LLM生成更diversity的结果。

\noindent\textbf{LLM settings. }
\jh{For LLM tasks, we utilize the state-of-the-art model gpt-3.5-turbo-0613 developed by OpenAI~\cite{gpt}.}
% 为了选择模型的temperature以达到最好的效果，我们设计了一个temperature实验对比了不同temperature下各任务的效果，并确定了各任务效果最好时的temperature，实验细节见附录XX
\jh{We conducted a temperature experiment to select the optimal temperature for maximizing the performance of the model in each task. Detailed information about this experiment is provided in Appendix~\ref{sec:temp-selection}.}
For raw APSRs generation, right code generation, violation code generation, and APSRs refinement tasks, the temperature of LLM is set to 0, 1, 0, and 1 respectively.
% 其余的所有参数都被设为默认值。使用zero-shot的方法设计prompt并限制max conversation length为5
All other model parameters are set to default, and we use the zero-shot approach to design prompts.

\delete{\noindent\textbf{Prompt Design. }
The complete prompts for raw APSRs generation, right code generation, violation code generation are shown in Figure~\ref{fig:cprompt-rule}, Figure~\ref{fig:cprompt-right}, Figure~\ref{fig:cprompt-violation} in appendix, respectively. The prompts of APSRs Refinement are shown in Figure~\ref{fig:cprompt-refine-one} (one $C_v$ in a group) and Figure~\ref{fig:cprompt-refine-multi} (multiple $C_v$ in a group) in appendix.}

\noindent\textbf{Preprocess.}
\change{
% 我们首先使用python 爬虫从库的官方文档中爬取了库的API列表.
We first crawl the API lists from the libraries' official websites.
}
We then use Tree-sitter~\cite{treesitter} to parse the Abstract Syntax Trees (ASTs) of the code. This allowed us to extract API source code from the library's source code. Furthermore, for each library, we manually prepared the required header files, necessary files for the APIs, and compilation options in advance \change{to help LLM generate the API calling code}. 
These are one-time tasks for a library and taking less than an hour for completion by an individual.

\noindent\textbf{APSRs Generation.}
\label{section-imple-validation}
When generating violation code, we use Tree-sitter~\cite{treesitter} to analyze the ASTs of code. 
While executing the code, we use two monitoring tools, ASAN~\cite{asan} and Valgrind~\cite{valgrind}, to capture memory-related runtime errors. \jht{We chose these tools because they are highly popular and provide comprehensive monitoring of potential issues during execution.} The maximum number of automated program repair attempts is set at 10 for Right Code Generation and 5 for Violation Code Generation. \change{All the complete prompts used by \toolname{} are available online \footnote{\url{https://github.com/icy17/GPTAid/tree/main/prompt}}}.
%在图XX的例子中, 15行对应的代码是sqlite3_open, 也就说明异常大概率是该API引起的。在确定了触发点后, 接下来需要确定该异常是否是由库引起的, 因为仍然存在其他因素引起异常的可能性, 比如图xx中, 示例代码的异常是由strlen抛出的, 该函数与目标API在同一行, 难以通过异常触发点所在行确定该问题。因此接下来我们会对异常调用路径进行分析, 通过判断异常路径是否存在目标API来确定异常是否是目标API导致的。

\noindent\textbf{\change{API Misuse Detection.}}
\change{
% 我们对\toolname生成的APSR进行聚类,并且确定了五个frequently occur的APSR pattern, 这五个pattern可以用来检测:memory leak, NULL pointer dereference, double free, use after free,以及use of uninitialize这些安全问题
We cluster the APSRs generated by \toolname{} and generate detection code templates for five frequently occurring patterns, including \textit{[API-A]} must not be called before \textit{[API-B]}, \textit{[API-A]} must be called after \textit{[API-B]}, the caller must not pass \textit{[VALUE]}, the parameter must not be used later, and parameter must be initialized. By applying these APSRs, \toolname{} can detect bugs such as memory leak, NULL pointer dereference, double free, and so on. To ensure efficiency, we employ intra-procedural analysis for detection.
}

% 接下来我将介绍不同类别的检测方案。
% \todo{add details}

%%%%%%%%%%%%%%%%IGNORE%%%%%%%%%%%%%%%%%%%%%%%%%%%%%%%
\ignore{n this stage, there are three tasks that require the use of LLM: right code generation, violation code generation, and automated program repair. In order to accomplish these tasks, LLM needs to utilize knowledge about the API as well as knowledge about compiling code, so we set the temperature to 1 for each of these tasks in order to make LLM generate as diverse results as possible. }
\ignore{
We started by manually categorizing and thoroughly analyzing the generated APSRs to build the initial detection template. \ignore{This process revealed \todo{XXX} distinct rule categories that are relevant to 304 APIs.}
% APSR类别信息见表XX
Following that, we crafted some manual detection code templates designed for identifying different types of misuses. These templates were used to automatically generate the final detection code. We use CodeQL~\cite{codeql} to find API misuses according to the detection code.}
\section{Evaluation}
\label{sec: evaluation}
% 实验设置, 运行环境配置
% 920-expr是最后的结果(output/auto-run-921-parse-rule-1)
% 1. effectiveness 分析规则总的FP FN, bug
    % end to end: bug检测 rule怎么获取 topXX
    % 各模块效果证明对LLM的调整是有效的
        % orig LLM
        % LLM + callgraph info + 多次
        % 2 + filter(no code check)
        % 3 + code check
% 对比实验 advance APISAN。
    % Advance对比 rule+bug
    % APISAN对比bug APEx
% 2. ablation study
% 3. 不同大模型效果: local-vicuna gpt-3. 5 gpt-4(部分）

% libxml2 sqlite3 libpcap openssl

In this section, we evaluate the effectiveness of \toolname{} in APSRs generation, API misuse detection, and individual components. We also compare \toolname{} with state-of-the-art approaches~\cite{Lyugoshawk2022, lv2020rtfm, liuIPPO2021}. Subsequently, we conduct an ablation study to show \toolname{}'s improvement on LLM, followed by an empirical analysis of the results and a case study.

\subsection{Settings}
% Table generated by Excel2LaTeX from sheet 'Sheet1'
\noindent\textbf{Dataset.}
% 为了评估toolname的effectiveness，我们基于github上库的star数以及在ubuntu上使用这些库的软件数，选择了真实世界中被广泛使用的4个popular的库：xxx。它们有着不同的功能，这些库的详细信息见表：
we utilized \change{five} datasets to evaluate the effectiveness of \toolname{}:
% \captionsetup[table]{labelfont={color=blue}}
% \arrayrulecolor{blue}
\begin{table}[!t]
  \centering
  % \vspace{-6pt}
  \small
  \caption{\change{Library Details}}
    \begin{tabular}{>{\centering\arraybackslash}m{6em}>{\centering\arraybackslash}m{10em}>{\centering\arraybackslash}m{6em}}
    \toprule
    \change{Library} & \change{Functionality} & \change{\#API} \\
    \midrule
    \change{libpcap} & \change{Network} & \change{71} \\
    \change{libxml2} & \change{XML parser} & \change{1614} \\
    \change{sqlite3} & \change{Database} & \change{294} \\
    \change{openssl} & \change{Cryptography} & \change{5478} \\
    \change{libevent} & \change{Event handling} & \change{401} \\
    \change{libzip} & \change{File compression} & \change{120} \\
    \change{zlib}  & \change{Data compression} & \change{69} \\
    \change{libcurl} & \change{Network Transfer} & \change{76} \\
    \midrule
    \change{\textbf{Total}} & \change{\textbf{/}} & \change{\textbf{8123}} \\
    \bottomrule
    \end{tabular}%
  \label{tab:lib_info}%
  % \vspace{-10pt}
\end{table}%
\captionsetup[table]{labelfont={color=black}}
\arrayrulecolor{black}

\vspace{2pt}\noindent$\bullet$\space\textit{Corpora of library source code ($C_{code}$).}
We selected eight widely-used libraries from different categories, including OpenSSL~\cite{openssl}, SQLite3~\cite{sqlite3doc}, libpcap~\cite{libpcap}, libxml2~\cite{libxml2}, libevent~\cite{libevent}, libzip~\cite{libzip}, zlib~\cite{zlib} 
and libcurl~\cite{libcurl}, based on their popularity on GitHub (measured by the number of stars) and their prevalence on Ubuntu. In total, we collected 8,123 APIs with 2.53M lines of code. Detailed information about these libraries is provided in \change{Table~\ref{tab:lib_info}.}  \delete{All of these libraries are used for API misuse detection.}
\\
\vspace {2pt}\noindent$\bullet$\space\textit{Ground-Truth dataset for APSRs Generation ($D_{gt}$).}
To evaluate the effectiveness of APSRs Generation, \change{we constructed a dataset by randomly selecting 25 APIs from \change{all the APIs across} each of the 8 libraries forming a total of 200 APIs,} \change{which intends to reduce the potential bias from frequently called APIs.}
\delete{The complexity of the API function call relationships made this a time-consuming task.}
% 我们首先按照advance和goshawk的方法生成并验证得到了部分的规则，为了防止漏掉规则，我们额外分析了API的代码，按照“什么样的参数会导致该API执行导致runtime error”这一规则分析了API的源码，并构造能够限制这样非法参数出现的规则，最终生成了376条APSR。
To create a comprehensive GroundTruth (GT), we first verified the APSRs generated by Advance~\cite{lv2020rtfm} and Goshawk~\cite{Lyugoshawk2022}, then analyzed the API source code and documentations to identify overlooked APSRs. Through our analysis, we generated \change{404} APSRs for 200 APIs in total.
% and manually generated 376 APSRs for these APIs.
\\
\vspace {2pt}\noindent$\bullet$\space\textit{\change{Comparison dataset ($D_{comp}$).}}
\change{To compare the effectiveness of \toolname{} with previous studies, we constructed a dataset consisting of bugs and their APSRs. The bugs include those were sourced from the previous studies within our scope and those were detected by \toolname{}. Since \toolname{} is implemented in user space, we excluded Linux kernel bugs, identifying 86 bugs from Advance and 10 bugs from Goshawk. Since IPPO~\cite{liuIPPO2021} does not disclose the locations of bugs, we only used the results from \toolname{}, Goshawk~\cite{Lyugoshawk2022}and Advance~\cite{lv2020rtfm}, which contained a total of 306 bugs and 58 APSRs.}
\\
\vspace {2pt}\noindent$\bullet$\space\textit{\change{Standard Dataset for API misuse detection (APIMU4C~\cite{gu2019apimu4c}).}}
% APIMU4C是最近的一个API misuse数据集，我们在这个数据集上分析\toolname{}在已知bug上的效果。APIMU4C中包含12个within our scope的bug
\change{APIMU4C is a standard dataset that focuses on API misuse. We utilized 12 bugs within our scope to evaluate the performance of \toolname{} in detecting bugs. }
\\
% detection：为了评估安全规则用来检测误用的效果, 我们根据github的star数量, 对每个库选择了10个较为常用的软件（共XX个软件）。
\vspace {2pt}\noindent$\bullet$\space\textit{Applications for API misuse Detection ($D_{app}$).}
To evaluate the effectiveness on API misuse detection, we selected 10 popular applications for each library based on the popularity (reflected by the number of github stars) to form a total of 47 applications. 
% 所有的软件github star的数量都大于1k，这也意味着使用这些软件的广泛使用.具体的软件信息见表XX。
All the applications have GitHub stars exceeding 1,000, indicating their widespread usage (details are shown in Table~\ref{tab:app_detail} in the Appendix).

\noindent\textbf{Platform.}
% . 5
We conducted experiments on a 64-bit server running Ubuntu 18.04 with 16 cores (Intel (R) Xeon (R) CPU v4 @ 2.10GHz), 440GB memory, and 11TB hard drive. 
\subsection{Effectiveness}
\label{sec:effectiveness}
In this section, we evaluate the effectiveness of \toolname{} on APSRs generation and API misuse detection.

\noindent\textbf{Effectiveness of APSRs Generation.}
\begin{table}[!t]
  \centering
    \vspace{-6pt}
  \small
  \caption{Effectiveness of APSRs Generation}
    \begin{tabular}{ccccc}
    \toprule
    Library & Precision & Recall & F1    & Cost Per API (\$) \\
    \midrule
    libpcap & 0.96  & 0.78  & 0.86 & 0.1 \\
    libxml2 & 0.89  & 0.68  & 0.77 & 0.12 \\
    sqlite3 & 0.91  & 0.82  &  0.86 & 0.1 \\
    openssl & 0.94  & 0.68  & 0.79 & 0.13 \\
    libevent & 1.00  & 0.63  &  0.77 & 0.11 \\
    libzip & 0.86  & 0.73  &  0.79 & 0.13 \\
    libcurl & 0.95  & 0.64  & 0.76 & 0.12 \\
    zlib & 0.88  & 0.73  & 0.80  & 0.11 \\
    \midrule
    Overall & 0.92  & 0.71  &  0.80 & 0.12 \\
    \bottomrule
    \end{tabular}%
  \label{tab:effective}%
\end{table}%
\change{We evaluated the effectiveness of \toolname{} in generating APSRs on $D_{gt}$ by manually analyzing their correctness.} We used three metrics to show the results: precision, recall, and F1 score.
 With a cost of only \$0.12 per API, \toolname{} generates 311 APSRs with a precision of \ablationFinalPre{} and a recall of \ablationFinalRecall{}, as shown in Table~\ref{tab:effective}.

% Table generated by Excel2LaTeX from sheet 'Sheet1'
% 误报：

% \paragraph{False Positives of APSRs Generation.}
%%%%%%%%%%%%%%%%%%%
\begin{figure}[t]
    \centering
    \includegraphics[width=0.9\linewidth]{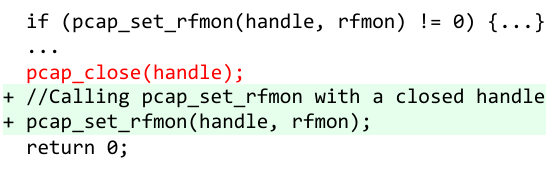}
    \caption{Example of a false positive in APSRs Generation}
    \label{fig:fp-example}
    % \vspace{-6pt}
\end{figure}
%%%%%%%%%%%%%%%%%%%

We analyzed these 24 false positives (FPs) and identified a primary cause: 75\% resulted from incorrect key operation identification during APSRs refinement. In this stage, \jh{LLM needs to identify the key operations by comparing correct and violation codes and generate the APSRs related to the key operations.} However, in cases with intricate modifications, the LLM often misidentifies key operations, leading to false positives. 
For example, Figure~\ref{fig:fp-example} illustrates an incorrect APSR for \texttt{pcap\_set\_rfmon} generated by LLM. 
The figure highlights differences between correct and violation code, with lines starting with + indicating added code. The violation code passes a closed variable to \texttt{pcap\_set\_rfmon}, causing the target API to use the variable after it has been closed, leading to a crash. Consequently, the correct rule should be: \textit{``Parameter 1 must not be closed before calling \texttt{pcap\_set\_rfmon}''}. However, LLM misinterprets the key operation as a violation code calling the API twice, and generates the APSR as \textit{``Parameter 1 (\texttt{p}) should not be called with the \texttt{pcap\_set\_rfmon} function more than once''}, thereby generate an incorrect APSR.

% 漏报：
% \paragraph{False Negatives of APSRs Generation.}
\label{sec:fn}
We conducted an analysis of these 117 false negatives (FNs) and identified three primary underlying reasons. 
\ding{182} \textbf{Missing in Raw APSRs Generation} accounts for 49.6\% of all FNs. Since \toolname{} relies on LLM to generate APSRs, and if LLM fails to generate the APSRs during the initial step, it leads to APSRs being missing. 
\ding{183} \dels{LLM’s unexpected behavior leads to missing.}\textbf{LLM's unexpected behavior}, contributing to 13.7\% of all FNs.
% 在生成violation code时，即使\toolname{}perform consistency check between modification and raw APSRs, 还是会有一些非预期的错误发生。比如modification可能在正确的位置修改了正确的参数相关变量，但是修改的方式是错误的，由于代码分析的复杂性，识别这种错误是非常困难的。这种错误会导致LLM不违反正确的APSR，不会导致runtime error，因此在分析执行结果时会错误地将该APSR分类为错误APSR然后过滤掉，导致了missing。另外，与第一点原因类似，在APSR refinement时，LLM会定位错误的key operation从而生成错误的APSR，这也会导致原本正确的APSR被漏掉。
When generating violation code, despite \toolname{} conducting a consistency check to identify the incorrect modifications, few unintended errors may go undetected, causing incorrect modifications. For example, a modification may modify the correct parameter in the correct location but with incorrect content. Identifying such errors becomes challenging due to the complexity of code analysis. 
The incorrect modifications cause LLM to deviate from violating the correct APSR, and generate a violation code without runtime errors. Consequently, \toolname{} mistakenly marks this APSR as incorrect, causing false negatives. Additionally, similar to the reason for FPs mentioned earlier, the LLM may misidentify key operations, generating incorrect APSRs and missing the originally correct ones.
\ding{184} \textbf{Failure to perform APSRs validation}. 
 Failure to generate the right code to invoke the API was responsible for \change{13.7\%} of all FNs. Since we validate APSRs by executing the code, any inability to generate the right code for API calls prevents validation, resulting in all APSRs for that API being missed. By analyzing the results, we found that  the success rate for generating right code is \rightSuccessRate{}. However, for the remaining 6.5\% of APIs, all APSRs are missing. 

\noindent\textbf{Analysis of generated APSRs. }
\label{sec:apsr_type}
\begin{table}[!t]
  \centering
    \vspace{-6pt}
  \small
  \caption{APSRs type}
  \begin{threeparttable}
    \begin{tabular}{ccccc|cc|>{\centering\arraybackslash}m{6em}}
    \toprule
    \multicolumn{5}{c|}{Value}            & \multicolumn{2}{c|}{Action} & \multirow{2}[0]{*}{Others} \\
\cline{1-7}    $C_{1}$* & $C_{2}$  & $C_{3}$ & $C_{4}$ & $C_{5}$ & $C_{6}$    & $C_{7}$ &  \\
    \hline
    \change{17}    & \change{90}    & \change{3}     & \change{7}     & \change{9}    & \change{85}    & \change{92}   & \change{8} \\
    \bottomrule
    \end{tabular}%
    \begin{tablenotes}
    \footnotesize
    \item[1]  $C_{n}$ denotes category-n, detailed in Section~\ref{sec:effectiveness}.
\end{tablenotes}
    \end{threeparttable}
  \label{tab:rule-type}%
  \vspace{-10pt}
\end{table}%
  % \vspace{-2pt}
% 我们参考了已有工作对APSR的分类方式，分析了\toolname{}在Dgt上生成的所有的APSR并将它们分为了八类，结果见表XX，除了已有工作中发现的五类关于参数值的APSR之外，\toolname还额外生成了两类action相关的APSR。
\jht{
We utilized the classification approach for APSRs from prior work~\cite{ZhouAnalysing2017, Zhongempir2020} to analyze and categorize all the APSRs generated by \toolname{} on $D_{gt}$ into eight categories, as detailed in Table~\ref{tab:rule-type}. This includes five previously identified categories and two new action-related types. These eight categories of APSRs are as follows:}

\noindent\jht{\ding{182} \textbf{Range. }This category defines constraints to avoid invalid ranges of parameter values. For example, a simplified APSR for the API \texttt{adler32} is \textit{``len must not be a negative value''}.}\\
\jht{\ding{183} \textbf{NULL.} This category specifies constraints to ensure parameter values are not NULL. For example, an APSR of \texttt{sqlite3\_open} is \textit{``ppDb parameter MUST NOT be null''.}}
\jht{\ding{184} \textbf{Member.} This category specifies constraints for values of parameter member. For example, an APSR of \texttt{pcap\_dump} is \textit{``fields of the struct must not all be -1''.} Unlike previous findings of APSRs in Java, C APIs' protected structs limit access to internal variables, resulting in fewer member APSRs.}
\\
\noindent\jht{\ding{185} \textbf{Relation.} This category defines the constraints on the relationships between different parameters. For example, an APSR of \texttt{sqlite3\_randomness} is \textit{``pBuf must be allocated with a size equal to or greater than N''.}}
\\
\noindent\jht{\ding{186} \textbf{Format.} This category describes constraints on parameter types or formats of content. For example, an APSR of API \texttt{sqlite3\_open} is \textit{``the filename must be null-terminated''.}}
\\
\noindent\jht{\ding{187} \textbf{Action-Do.} This category of rules describes constraints on the actions required for the parameters. For example, an APSR of \texttt{zip\_register\_progress\_callback} is \textit{``The `za` should be initialized by calling zip\_open''.}}
\\
\noindent\jht{\ding{188} \textbf{Action-Not Do.} This category specifies actions that are disallowed for parameters. For example, an APSR of \texttt{curl\_share\_cleanup} is \textit{``must not be freed before''.}
\\
\ding{189} \textbf{Others.} This category includes APSRs that don't fit into the above categories. For example, an APSR of \texttt{gzflush} is \textit{``The caller MUST NOT use \texttt{gzflush} with the flush parameter if there is a seek request pending''.}
}
\\
\noindent\textbf{Analysis of detected API misuse. }
\begin{table}[!t]
  \centering
    \vspace{-6pt}
  \small
  \caption{Detection results of \toolname{}}
    \begin{tabular}{>{\centering\arraybackslash}m{6em}>{\centering\arraybackslash}m{4em}>{\centering\arraybackslash}m{6em}>{\centering\arraybackslash}m{4em}}
    \toprule
    Library & \#APIs & \#APSRs & \#Bugs \\
    \midrule
    libpcap & 17    & 36    & \change{7} \\
    libxml2 & 30    & 19    & 5 \\
    sqlite3 & 94    & 136   & \change{7} \\
    openssl & 163   & 206   & \change{51} \\
    libevent &  47  &  48  & \change{140} \\
    libzip &  27  &  49  & 0 \\ 
    zlib &  22 &  43  & 0 \\ 
    libcurl &  30  &  42  & 0 \\ 
    \midrule
    Total & 431   & 579   & \change{210} \\
    \bottomrule
    \end{tabular}%
    % \vspace{-10pt}
  \label{tab:misuse_re}%
\end{table}%
We evaluated the effectiveness of \toolname{} in API misuse detection using APSRs generated by \toolname{}. 
% 我们首先分析了GPTAid在APIMu4C数据集上的效果。我们发现在12个within our scope的bug中，gptaid能够检测出11个bug，这些bug是已有工作很难检测出来的cite。我们分析了missed bug，然后发现这是一个missing unlock bug，由于GPTAid使用的monitor无法在动态执行过程中捕获该问题造成的异常，因此GPTAid无法生成对应的规则，从而导致了无法检测该bug.我们接下来分析了GPTAid在检测新bug方面的能力
\change{We first analyzed the effectiveness of \toolname{} on APIMU4C, where it successfully identified 11 out of 12 bugs. The missed bug is a lock-missing-unlock bug. \toolname{} failed to detect it because \toolname{} could not capture exceptions caused by the missing unlock during execution, thus preventing the generation of the necessary APSRs for bug detection.} We then evaluated the ability of \toolname{} to detect new bugs. 
\change{To ensure cost-effectiveness}, we selected a subset of APIs that were called more than 10 times within $D_{app}$ for misuse detection, totaling 431 APIs. \toolname{} generated 579 APSRs for this subset. \change{Not every API has an APSR, as we did not apply additional filtering criteria.}
% \paragraph{API misuse Detection.}
% Table generated by Excel2LaTeX from sheet 'Sheet1'
We applied CodeQL to detect API misuse within $D_{app}$, detecting a total of \change{210} unknown bugs, \change{150} of which have been confirmed by the developers after reports. Results of API misuse detection are shown in Table~\ref{tab:misuse_re} and the details of API misuse are shown in Table~\ref{tab:bug-detail} in Appendix.
\change{
The precision of \toolname{} on bug detection is 77.2\%, which is acceptable for static analysis-based detection, and it is higher than that of comparable tools like IPPO. Our analysis shows no false positives (FPs) from incorrect APSRs, as we only generate detection code template for frequently occurring rules, minimizing the impact of errors. One primary cause of false positives (FPs), accounting for 45.2\% of all FPs, is the limitation of intra-procedural analysis. For example, \toolname{} detects potential null pointer dereference by verifying the presence of a NULL check before calling the API. However, if this check is located in another function, the intra-procedural analysis may erroneously report a bug due to the absence of the check within the current function.
}
\delete{We analyzed the security impact of these bugs and found that 173 of them result in NULL pointer dereference (CWE-476), which can lead to software crash. Additionally, 10 of these misuse cause Improper Resource Shutdown or Release (CWE-404), potentially resulting in Denial of Service (DoS). \jh{We use a fixed misuse as a case study to show a common security impact in Section~\ref{sec:case-study}. }}

% Table generated by Excel2LaTeX from sheet 'Sheet1'

%我们分析结果发现，XX没有在使用libxml2的软件中找到bug。为了分析原因，我们对libxml2的API的生成的规则进行了分析。我们发现libxml2的源码中有较全面的error handling代码，自身存在的APSR较少。详细来说，当API由于违反APSR而崩溃时意味着不符合要求的变量导致API内部代码出现错误，比如当API中存在解引用操作时，如果操作对象是用户传递的参数，那么当用户违反APSR传递了NULL指针时，API中的解引用操作会导致crash。而libxml2库的API代码中，在使用用户传递的参数前，普遍存在着对参数是否符合要求的检查，因此存在的APSR较少。

\ignore{We also analyzed the impact of the occurrence of false positives in APSRs on API misuse detection. Following the clustering of APSRs, we observed that the majority of false positives tend to be associated with infrequently occurring rules. Consequently, Manually analyzing these rules to filter out false positives is acceptable. \ignore{We filter false positives for \todo{XX APIs in XXX (time).}}}
\subsection{Evaluation of Individual Components}
\label{sec:individual}
% Table generated by Excel2LaTeX from sheet 'Sheet1'
% \todo{In this section, we evaluate the effectiveness of individual components(i.e., raw APSRs generation, APSRs validation and APSRs refinement). }

\noindent\textbf{Effectiveness of Raw APSRs Generation. }
We evaluated the effectiveness of raw APSRs generation on 200 APIs in $D_{gt}$, finding \toolname{} generated \change{2858} raw APSRs with a recall of \change{84.4\%}, as shown in Table~\ref{tab:individual}.
This stage relies on the LLM's capabilities, and the occurrence of false negatives (FNs) is also linked to these capabilities. We attempted to identify the causes of these FNs by examining the API source code.
One possible reason for FNs is that the LLM generates APSRs from source code analysis, focusing on the functionality of API parameters. Occasionally, this results in stereotypical inferences, contributing to FNs. For example, when a parameter serves to pass a file name, the LLM tends to generate rules focused on file names, such as: \textit{``the filename must be validated to ensure it does not contain any path traversal''}. However, stereotypical inferences may limit the LLM's code analysis, causing it to overlook other rules which is unrelated to the functionality. This accounts for \change{44\%} of all FNs.
%生成raw APSR的工作完全由LLM完成，该阶段的漏报与LLM的能力相关，我们无法得知准确的造成FN的原因。我们尝试通过分析漏报规则对应的API代码对这些漏报的原因进行分析。 通过分析，我们发现了与漏报比较相关一个原因。LLM在分析源码生成规则时，会根据该参数在API源码中的功能生成对应的规则，有的时候会陷入思维定势，这导致了FN。比如对于传递file name的参数，LLM倾向于生成一些与file name有关的规则比如：文件名必须合法，xxx等等，然而这会阻碍LLM对于代码的分析，导致LLM漏掉规则。这种原因导致的漏报占所有漏报的48.3%
% 比较有趣的是，FN并不是由于prompt中缺少信息导致的（dh free和Nconf free），并不是提供的信息足够多就能够尽可能激发LLM的知识，生成更多的规则。比如对于Nconf free和DH free来说，DH free有着更多的代码 （loc），这些代码包含了丰富的语义及代码特征，而Nconf free仅仅调用了indirect的函数。然而，LLM无法生成DH free相关的规则，但是对于Nconf free可以生成几乎完全的规则。我们随机生成了5次，结果都是相同的。

\noindent\textbf{Effectiveness of APSRs Validation. }
% 在这个部分，我们通过评估validation失败的概率（FR）以及validation成功但是结果出错的概率（FPR）来展现validation的有效性。通过人工分析validation的结果，\toolname{}在200个API的2513条规则上的正确率为XX，漏报率为xx。
In this section, we calculate the following: True Positives (TP), which are APSRs that are successfully and correctly validated; False Positives (FP), which are APSRs that are successfully but incorrectly validated; and False Negatives (FN), which are APSRs that failed to validate. We then use recall ($\frac{TP}{TP + FN}$) and precision ($\frac{TP}{TP + FP}$) to evaluate the effectiveness of \toolname{} in APSRs validation. Based on our analysis, \toolname{} achieved a precision of 96\% and a recall of \change{80.2\%} in APSRs validation.

\jht{
% 我们分析了实验结果并发现了导致validation失败的一个主要原因。llm在对正确代码进行修改时会引入难以被修正的错误，这个原因导致的FN占所有FN的xxx。比如存在一些规则约定：xxxthread，LLM为了生成违反规则的代码，需要生成多线程调用的代码，并保证多线程调用方式违反目标规则。在这种情况下LLM会难以生成通过编译并执行到目标API的violation code。
We analyzed all FNs and identified a major cause. 
The LLM struggles to generate right code and violation code, resulting in \change{75\%} FNs. 
For example, a raw APSR of \texttt{sqlite3\_bind\_int} is: \textit{``Parameter 3: Prevent the use of the `iValue` parameter in a multi-threaded environment without proper synchronization''.} To generate the violation code, LLM must generate a multi-threaded call that violates this specific rule, which is a complex task. LLM struggles to generate that complex violation code that can be compiled, leading to \toolname{}'s inability to validate this APSR.}

% 我们分析了FP，并发现了一个导致FP的主要原因。在生成violation code时llm对代码的修改可能是不正确的。这个原因占所有FP的52.6/%.当LLM试图违反某一规则时，可能会引入一些错误，或者并不是完全按照规则的要求违反规则。比如API xxx 的raw APSR为XXX，LLM在尝试违反该规则时，同时修改了parameter 1 的初始化方式，使parameter 1 变成了invalid的变量，这导致了violation code执行出现runtime error，然而该error与raw APSR是无关的，导致该APSR被错误地分类为正确的APSR。虽然我们对violation方式的正确性进行了coarse的check，但仍有极少的unexpected  behavior出现。
\jht{We analyzed the results and found a major cause of FPs: errors introduced by LLM during the modification. This accounts for \change{52\%} of all FPs. When LLM tries to violate an APSR, it may introduce errors or not fully comply with the APSR's requirements for violation. For example, a raw APSR of API \texttt{sqlite3\_vtab\_in\_frist} is \textit{``Parameter 1: The pVal parameter should not be NULL''.} When LLM tries to violate the rule, it inappropriately alters the initialization of the parameter 2, turning it into an invalid variable and causing a runtime error. This unrelated error leads to the APSR being mistakenly identified as correct. Although \toolname{} conducts coarse checks on violation modification correctness, a few unexpected behaviors still arise.}

\noindent\textbf{Effectiveness of APSRs Refinement. }
\begin{table}[!t]
\vspace{-6pt}
  \centering
  \small
  \caption{Effectiveness of Individual Components}
    \begin{tabular}{c|c|cc|c}
    \hline
    % \multicolumn{1}{p{8em}}{\tabincell{c}{SSL\_CTX\_set\_ssl \\ \_version}}
    \multirow{2}[4]{*}{Libs} & \multicolumn{1}{m{4.5em}|}{\tabincell{c}{raw APSRs\\generation
}} & \multicolumn{2}{m{4.96em}|}{\tabincell{c}{APSRs\\validation}} & \multicolumn{1}{m{4.5em}}{\tabincell{c}{APSRs\\refinement}} \\
\cline{2-5}          & Recall & Precision    & Recall    & Presicion \\
    \hline
    OpenSSL & \change{0.85}  & \change{0.94}  & \change{0.84}  & 0.93 \\
    \hline
    SQLite3 & \change{0.89}  & 0.96  & \change{0.82}  & 0.93 \\
    \hline
    libxml2 & \change{0.84}  & \change{0.95}  & \change{0.81}  & 0.88 \\
    \hline
    libpcap & \change{0.95}  & \change{0.95}  & 0.87  & 0.94 \\
    \hline
    libevent & \change{0.84}  & \change{0.97}  & \change{0.69}  & 0.9 \\
    \hline
    libzip & \change{0.82}  & \change{0.96}  & \change{0.78}  & 0.88 \\
    \hline
    zlib  & \change{0.77}  & \change{0.98}  & 0.88  & 0.91 \\
    \hline
    libcurl & 0.78  & 0.98  & 0.74  & 0.94 \\
    \hline
    \textbf{Total} & \textbf{\change{0.84}}  & \textbf{0.96}  & \textbf{\change{0.80}}  & \textbf{0.91} \\
    \hline
    \end{tabular}%
  % \vspace{-12pt}
  \label{tab:individual}%
\end{table}%
% 在这一部分，我们通过计算能够正确转化成下游检测规则的APSR占raw APSR的比例来评估refinement模块的效果。我们将模糊的规则或是refinement与原violation无关的规则都认为是错误的refinement。该模块生成了XXX条正确的规则，pre为90/%.
In this section, we assess the effectiveness of APSRs refinement by measuring the proportion of APSRs that are accurately transformed into downstream rules (precision). We categorize refinements as incorrect if they result in general rules or deviate from the original violation. Based on our analysis, \toolname{} achieves an accuracy of 91.5\%.
% 我们分析了错误的结果，发现了一个导致错误的主要原因。LLM会识别错误的modification operation。对于极少数具有复杂的modification操作的APSR来说，LLM难以从多个代码中定位到正确的modification operation，因此基于错误的operation总结出错误的APSR。比如对于多线程操作，modification除了导致error的操作外还包含了大量的多线程相关的操作，LLM在分析这种复杂的代码时会生成的与多线程相关的错误结果。
\jht{
We analyzed the incorrect refinement results and identified a primary error source:  LLM identifies incorrect key modification operations. In few cases of violation code with complex modifications, LLM struggles to select the key operation from among multiple code snippets, resulting in inaccurate APSRs summaries. For example, in the context of multi-threaded APSRs, modifications include numerous multi-threading-related operations and key operations that cause runtime errors. LLM generates incorrect multi-threading-related APSR based on these operations.
}

\subsection{Compare to the State-of-the-Art}
\label{sec:compare-study}
\delete{In this section, we compare the effectiveness of \toolname{} with three other tools: Advance~\cite{lv2020rtfm}, IPPO~\cite{liuIPPO2021}, and Goshawk~\cite{Lyugoshawk2022} for APSRs generation and API misuse detection.}
% 我们从section I总结的每一类中各选择了一个frequently referenced in previous studies的工具。
\change{In this section, we compared the effectiveness of \toolname{} with three state-of-the-art tools: Advance~\cite{lv2020rtfm}, IPPO~\cite{liuIPPO2021}, and Goshawk~\cite{Lyugoshawk2022}. These tools were selected from the categories summarized in Section~\ref{sec:intro} (analyzing API calling code, documentation, and API source code) and were frequently referenced in previous studies.}
 \delete{Advance is a documentation-based approach can generate APSRs and perform API misuse detection, IPPO detects bugs in a software based on inconsistent security operations of path-pairs in the client-code, and Goshawk is based on the library source code, identifies functions that perform memory release/allocation, and detects memory-related bugs resulting from their misuse.}
 % 我们在D_comp上评估这三个工具和GPTAid在APSR generation和bug detection这两个方面的效果，结果展示在xxx.由于Advance没有公开完整的源码，因此我们仅仅从生成的APSR数量和检测到的bug数量两个方面评估效果。
 \change{We evaluated the effectiveness of these three tools and \toolname{} on $D_{comp}$ for APSRs generation and API misuse detection, based on the number of APSRs generated and bugs detected.}
\delete{Advance and IPPO not only generate APSRs but also generate return value-related security rules and detect other types of bugs. To compare the effectiveness of these tools and \toolname{}, we only calculate the APSRs generated by them for comparison.}
The results of the comparative experiments are shown in Table~\ref{tab:compare}.

% Table generated by Excel2LaTeX from sheet '对比实验'
\delete{\begin{table}[!t]
\small
  \centering
  \caption{Comparison with State-of-the-Art tools.}
  \begin{threeparttable}
    \begin{tabular}{ccccc}
    \toprule
          & \toolname{} & Advance & IPPO  & Goshawk \\
    \midrule
    APSRs precision & 0.92    & /    & /    & 1 \\
     APSRs recall & 0.71    & 0.08 & /    & 0.04 \\
    \#Bugs & 183    & 8    & 0    & 0 \\
    \bottomrule
    \end{tabular}%
    \begin{tablenotes}
    \footnotesize
    \item[1]  / means we cannot calculate this result.
\end{tablenotes}
\end{threeparttable}
  \label{tab:compare}%
  % \vspace{-6pt}
\end{table}%
}
\begin{table}[!t]
  \centering
  \vspace{-6pt}
  % \arrayrulecolor{blue}
    \small
  \caption{Comparison with State-of-the-Art tools.}
  \begin{threeparttable}
    \begin{tabular}{cccccc}
    \toprule
          & \change{\toolname{}} & \change{Advance} & \change{IPPO}  & \change{Goshawk} & \change{$D_{comp}$} \\
    \midrule
    \change{\#APSRs} & \change{53}    & \change{7}    & \change{/}    & \change{8} & \change{58}\\
    \change{\#Bugs} & \change{243}    & \change{99}    & \change{0}    & \change{10} & \change{306}\\
    \bottomrule
    \end{tabular}%
    \begin{tablenotes}
    \footnotesize
    \item[1]  / means we cannot calculate this result.
\end{tablenotes}
\end{threeparttable}
  \label{tab:compare}%
  \vspace{-13pt}
\end{table}%

\noindent\textbf{Advance.}
% advance通过识别文档中的强烈语气来提取文档中的APSR并用APSR来检测API误用。jh
Advance extracts APSRs from documentation by identifying strong sentiments in documents and utilizes these APSRs to detect API misuse.
We compared its effectiveness with \toolname{} in both APSRs generation and API misuse detection. 
We consider all APSRs described in the documentation as those generated by Advance and count the associated API misuses as bugs found by Advance.
We evaluated Advance on \change{$D_{comp}$} and found that Advance can extract a maximum of \change{7} APSRs and identify up to \change{99} API misuse. 
In comparison, \toolname{} generated \change{6} times more APSRs. 
% 为了探寻advance效果不理想的原因，我们对advance的结果进行了深入分析，并发现了两个可能的原因。
To understand the unexpectedly results of Advance, we thoroughly analyzed the results and documentations and identified two possible reasons for the missing APSRs.
\ding{182} Lack of Explicit Description in API Documentations. 
Through our analysis of the API documentations, we found that \change{88.0\%} of the APSRs in \change{$D_{comp}$} are not explicitly written in the documentation. These rules are often considered as common knowledge among developers and are therefore omitted.
% Most content of documentation emphasizes API functionality, sometimes overlooking descriptions of common-sense rules for conciseness. 
\ding{183} Descriptions without Strong Sentiment.
% 由于advance通过识别文档中的强烈语气提取IA，如果描述APSR的语句没有强烈语气，那么advance没办法从文档中提取相应的描述。这些原因导致了Advance缺失了大量APSR。
Advance extracts IAs by recognizing strong sentiment in documentations. However, without strong sentiment in APSR descriptions, Advance cannot extract corresponding information, leading to missed APSRs. 
\change{
% 在advance的结果中，toolname能生成xx条APSR检测到xx个bug。miss的原因主要是toolname使用的monitor无法检测info leak这一问题，导致漏掉对应的bug和APSR。
\toolname{} missed 2 APSRs and 58 bugs detected by Advance because its monitors couldn't catch the exceptions during dynamic execution. Despite this, \toolname{} identified more APSRs and bugs, showing that it performs better than the documentation-based method.
}

\noindent\textbf{IPPO.}
IPPO identifies bugs by identifying inconsistent security operations within path-pairs in the API calling code. We applied IPPO on $D_{comp}$ for bug detection and compared the results with those obtained using \toolname{}. The results showed that IPPO did not detect any bugs.
By analyzing the results, we identified two main reasons for the missing bugs:
\ding{182} IPPO faces challenges in identifying security operations. IPPO detects bugs by comparing path pairs for inconsistent security operations but struggles to determine if an API is a security operation due to its limited understanding of APIs. This limitation can result in missed bugs.
\ding{183} IPPO cannot detect bugs where inconsistent security operations are absent. IPPO identifies bugs by finding inconsistent security operations in path pairs. However, if API misuse occurs due to the absence of a security operation in all paths, IPPO cannot detect it because there is no inconsistency.

\noindent\textbf{Goshawk.}
Goshawk can identify APIs in libraries that have allocation and deallocation functionality and detect use-after-free and double-free bugs based on the identified APIs. Goshawk can extract \change{8} APSRs and detect up to \change{10} bugs on \change{$D_{comp}$}, \change{compared to the 53 APSRs and 243 bugs identified by \toolname{}}.
By analyzing the results, we identified two primary reasons for the FNs in Goshawk:
\ding{182} Focusing on the allocation/deallocation presents limitations. As previously mentioned, Goshawk's rule extraction from library code is constrained by predefined analysis rules, preventing it from extracting other types of rules. Consequently, Goshawk can only extract allocation/deallocation APSRs and cannot detect other types of bugs. Allocation/deallocation-related APSRs comprise only \change{25.9\%} of the APSRs in \change{$D_{comp}$.} 
\ding{183} \jh{Filtering functions based on name analysis results in missing.} Goshawk leverages Natural Language Processing (NLP) techniques to analyze function names before conducting static analysis on library functions. This process filters out functions whose names lack relevance to allocation/deallocation operations, which may incorrectly exclude relevant APIs with less obvious names, leading to false negatives.
\change{
% 在goshawk的结果中，toolname能生成xx条APSR检测到xx个bug。miss的原因主要是toolname只进行了intra分析，对于需要inter分析的xx个bug会被漏掉。
\toolname{} missed 3 APSRs and 5 bugs detected by Goshawk. The missed APSRs are caused by ``Missing in Raw APSRs Generation'' and ``Failure to perform validation'' (details in Section~\ref{sec:effectiveness}). The missed bugs primarily occurred because \toolname{} only performs intra-procedural analysis, missing bugs that require inter-procedural analysis. The results show that \toolname{} generates more diverse APSRs and detects more bugs than Goshawk.
}

\subsection{Ablation Study}
\label{sec:ablation}
% Table generated by Excel2LaTeX from sheet 'Sheet1'
\begin{table}[!t]
\arrayrulecolor{black}
\vspace{-6pt}
  \centering
  \small
  \caption{Results of ablation study}
  \label{tab:ablation}%
  \begin{threeparttable}

    \begin{tabular}{>{\centering\arraybackslash}m{9em}cccc}
    \toprule
          & FP    & TP    & Precision & F1 Score\\
    \midrule
    LLM & \change{2517}  & \change{341}   & 0.12  & 0.21 \\
    LLM+$S_v$ & \change{74}    & \change{284}   & 0.79  & 0.74 \\
    LLM+$S_v$+$S_r$ &  24 & \change{287}   & 0.92  & 0.80 \\
    \bottomrule
    \end{tabular}%
    \begin{tablenotes}
    \footnotesize
    \item[1] LLM means directly using LLM to generate APSRs.
    \item[2] $S_v$ is APSRs Validation. $S_r$ is APSRs Refinement.
\end{tablenotes}
  \vspace{-10pt}
    \end{threeparttable}
\end{table}%

In this section, we evaluated the contribution of \toolname{}  to the enhancement of LLM. We evaluated the effectiveness of directly using LLM, LLM+APSRs validation (LLM+$S_v$), and LLM+APSRs validation+APSRs refinement (LLM+$S_v$+$S_r$). We refer to the APSRs generated directly using LLM as $APSR_1$, the APSRs generated by LLM+$S_v$ as $APSR_2$, and the APSRs obtained by LLM+$S_v$+$S_r$ as $APSR_3$. We calculate precision and F1 score as evaluation metrics. 

\noindent\textbf{Contribution of APSRs Validation.}
In this section, we evaluated the impact of APSRs validation on enhancing LLM accuracy. We conducted a comparison between the $APSR_1$ obtained directly using LLM and the $APSR_2$ obtained after APSRs validation, and the results are presented in Table~\ref{tab:ablation}. The findings reveal a significant improvement in precision, increasing from 0.12 to 0.79 \jh{and the F1 score increasing from 0.21 to 0.74}. This improvement shows the effectiveness of our method in validating the correctness of APSRs.

\noindent\textbf{Contribution of APSRs Refinement.}
% 问题：这部分是否需要去掉生成规则时一起生成的violation example？
In this section, we evaluated the impact of APSRs refinement on generating concrete APSRs. We compared the $APSR_2$ with the $APSR_3$. The results are presented in Table~\ref{tab:ablation}. The results indicated an improvement in both precision and f1 score. Precision has increased from 0.79 to 0.92, and the F1 score has risen from 0.74 to 0.80. This suggests that the APSRs refinement stage generate concrete APSRs. 

% Table generated by Excel2LaTeX from sheet 'Sheet1'
% \begin{table}[!t]
%   \centering
%   \caption{Results of using GPT-4 on APSRs generation}
%     \begin{tabular}{ccc}
%     \toprule
%           & Precision & Recall \\
%     \midrule
%     GPT-4 & 0.21  & 0.67  \\
%     \textbf{GPTAid} & \textbf{0.92} & \textbf{0.71}  \\
%     \bottomrule
%     \end{tabular}%
%   \label{tab:gpt4-re}%
% \end{table}%

\subsection{Findings}

%从每一个case中, 总结一条finding, 不要纯描述一个case。
% 1. bug detection发现使用EVP_DigestInit_ex文档中存在问题, 因此很多使用该API的场景都出现了bug（数量
\noindent\textbf{Document errors leads to API misuse.}
% \begin{figure}
%     \centering
%     \includegraphics[width=0.7\linewidth]{Figure/finding.pdf}
%     \caption{Incorrect example code in the documentation}
%     \label{fig:find-example}
%     % \vspace{10pt}
% \end{figure}
We discovered that APIs whose code examples in documents are incorrect are more likely to result in API misuse. Based on our analysis of the results of API misuse detection, we observed that the \texttt{EVP\_DigestInit\_ex} API exhibited the highest percentage of misuse in the applications. After analyzing the documentation, we discovered that the API documentation~\cite{evpwrongdoc} includes example code for using the \texttt{EVP\_DigestInit\_ex} API. However, the example fails to check if the parameter 1 is NULL, posing a potential security risk of null pointer dereference. The API misuse in the documentation may influence users to adopt a similar usage when calling this API. 

\noindent\textbf{APSRs enhance the quality of documentation.}
% 对于304个API我们生成了XX条规则，补全了文档，提高了文档质量。通过分析Dgt数据集中的APSR是否在文档中有描述，我们发现XX的APSR是没有被明确描述的，而这导致了使用者无法从文档中了解到需要遵守的APSR从而更容易导致API误用。自动化分析文档的安全工具比如advance也无法从这种文档中提取可用的规则。我们的方法可以被用来自动化生成APSR，进而补充文档的安全描述，提升文档的质量。
We generated 579 APSRs for 431 APIs, improving the quality of their documentation. 
% 我们对APSR在GT上得到的结果进行了分析，发现其中XX的规则
We discovered that \change{61.3\%} of the APSRs we identified lacked explicit descriptions in the documentation. This makes it challenging for users to learn the APSRs they must follow from the documentation, consequently increasing the risk of API misuse. Existing approaches in rule extraction through documentation analysis, such as Advance~\cite{lv2020rtfm}, face limitations due to the lack of security descriptions in the documentation. 
Our approach can automate the generation of APSRs to enhance the security descriptions in the documentation. \change{We reported these missed APSRs to the developers of the libraries, and \confirmedAPSR{} of them were confirmed. Some developers considered these APSRs common sense and didn't include them in the documentation. However, without clear guidelines, developers may accidentally violate these APSRs, leading to security issues. Additionally, relying solely on common sense for bug detection can cause issues, as some APIs have internal checks. This means violations may not cause security issues, leading to false positives in bug detection.
}

% relying solely on common sense to detect misuse for all APIs can cause issues, as some APIs have internal checks (e.g., null checks) that prevent issues even with illegal parameters. In this case, using common sense for all APIs might lead to many false positives.}

\noindent\textbf{\change{Security impact.}}
\change{We analyze the security impact of bugs detected by \toolname{}, including those in $D_{comp}$ and APIMU4C, categorized using the CWE standard. These bugs can be classified into several types: NULL pointer dereference (CWE-476), double free (CWE-415), and improper resource shutdown or release (CWE-404).
These bugs can lead to serious security implications, such as information leaks, memory corruption, crash, code execution, and so on.}
We use a misuse as a case study to show a common security impact in Section~\ref{sec:case-study}. 

\subsection{Case study}
\label{sec:case-study}
% \paragraph{Null pointer dereference (CWE-476).}
\begin{figure}
    \centering
    \includegraphics[width=0.9\linewidth]{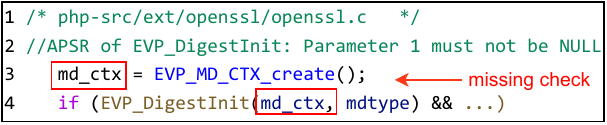}
    \caption{Example of null pointer dereference}
    % \vspace{-6pt}
    \label{fig:case-null}
\end{figure}
% php-src是source code of a popular general-purpose scripting language. \toolname{} reported an API misuse cause a NULL pointer dereference, leading to crash, the misuse code and APSR generated by \toolname{} are shown in XX. Specigfically, 
\jh{Php-src~\cite{php-src} (36.6k stars on GitHub) is the source code for a popular scripting language. \toolname{} found an API misuse in php-src causing a NULL pointer dereference, potentially leading to a crash. The code snippet causing misuse and the APSR generated by \toolname{} are shown in Figure~\ref{fig:case-null}.}
% \toolname{} 发现了一些空指针解引用问题，比如in XXX as shown in FigureXXX。md_ctx通过调用API EVP_MD_VTX_create获得（line2），而后该变量被直接传递给EVP_DigestInit而没有先检查该变量是否为NULL。在EVP_MD_CTX_create函数调用失败时会出现空指针解引用问题导致软件crash。
Specifically, 
 \jh{the APSR requires that parameter 1 of this API must not be NULL. Therefore, the caller of this API should check if parameter 1 is NULL before calling this API.}
 In the source code of php-src, the variable \texttt{md\_ctx} is obtained by calling \texttt{EVP\_MD\_CTX\_create} (line 2), and is then passed directly to  \texttt{EVP\_DigestInit} (line 3) without checking whether it is NULL first, thereby violating the APSR for \texttt{EVP\_DigestInit}. This results in a null pointer dereference when \texttt{EVP\_MD\_CTX\_create} fails, leading to a system crash.
However, this APSR is absent in the documentation, and the provided example of API usage is incorrect. Therefore, documentation-based approaches like Advance cannot generate this APSR, leading to the missing of this API misuse. Additionally, since \texttt{EVP\_DigestInit} is only called once, detecting misuse through comparing different API usage patterns in the API calling code is not possible.

\section{Discussion}
\label{sec:discuss}
% 在这一section，首先我们将介绍我们在使用LLM时设计prompt的一些经验。接着我们会介绍我们通过分析我们找到的API误用得出的避免API误用要注意的问题。最后我们将介绍本工作的limitation以及我们的future work。
In this section, \change{we begin by presenting our exploration of using GPT-4 for APSRs generation and using \toolname{} on new APIs.} We then share insights into prompt design with LLM, and lessons for preventing API misuse. \change{Finally, we summarize \toolname{}'s limitations and outline future work.}
\subsection{\change{Exploration on GPT-4}}
\change{
% GPT-4是一个
% 众所周知，GPT-4的效果好于GPT-3.5。我们探索直接使用GPT-4是否能够在生成APSR上达到一个比较好的效果。我们在Dgt上对GPT3，GPT4和GPTAid （with gpt-3.5）的效果进行了对比。GPT-4达到了recall XX，precision xx，结果见表xx。从结果可以看出GPT4的开销与GPTAid相似，然而效果远远低于GPTAid。这证明了GPTAid在生成APSR上的效果。我们分析GPT4的结果发现，GPT4比GPT-3.5更容易发生Section XX中提及的stereotypical inference问题，这也就导致GPT4漏掉更多的APSR。
GPT-4 is a powerful LLM that outperforms GPT-3.5.} \change{We explored whether using GPT-4 directly can achieve better results in generating APSRs. We provided the same prompt used for raw APSRs generation to GPT-4 and compared the results with \toolname{} (using GPT-3.5) on $D_{gt}$. GPT-4 achieves a recall of 0.67 and a precision of 0.21, compared to the 0.71 recall and 0.92 precision of \toolname{}. These results are significantly lower than those of GPTAid, demonstrating GPTAid's effectiveness in generating APSRs. Our analysis shows that GPT-4 is more prone to the stereotypical inference problem mentioned in Section~\ref{sec:individual} than GPT-3.5, leading to more missed APSRs.
}
\subsection{\change{Exploration on new APIs}}
% LLM在大量数据上进行训练，这些训练数据中可能包括API的使用代码，这可能导致LLM更容易能够生成对应的API调用代码。因此，我们探索了GPTAid在正确代码生成这一任务上，对于新API的代码生成效果。我们在这八个库上找到了10个2021.9月之后（after llm trained）新加入到库中的API，并使用GPTAid生成它们的API calling code。通过分析结果我们发现，成功生成了9个API，这证明了GPTAid具有生成未学习过的API调用代码的能力。仅有的那个失败的API有很多的参数，这意味着它依赖的前置条件难以满足，处于cost考虑，我们限制自动修复的次数为10，增加这个上限能够帮助GPTAid生成这种复杂API的调用代码。
% 我们分析了调用失败的API,发现正确调用该API需要构建某种特殊结构的文件输入,这种条件难以通过修改代码得到满足,该问题并不是由于新API引起的.这证明了\toolname{}在生成正确代码任务上的效果
\change{Since LLM is trained on a large amount of data that may include API usage code, we aimed to explore whether LLM relies on this data to generate correct API calling code. Considering the training data for gpt-3.5-turbo-0613~\cite{gpt3-5} is up to September 2021, we identified 10 new APIs from eight libraries added after that date, and used \toolname{} to generate their API calling code. Based on our analysis, we found that \toolname{} successfully generated code for 9 out of 10 new APIs. The API that \toolname{} failed to generate code for required a specific file structure as input, which is difficult to satisfy by only modifying the code. The issue was not due to the new API itself, demonstrating the effectiveness of \toolname{} on API calling code generation.}

% lesson of LLM
\subsection{Lessons for prompt design}
\label{sec:discuss-llm}

\noindent\textbf{Insufficient information in prompt leads to fabrication.} \change{We observed that LLM tends to generate fabricated results when there is a lack of information or clear instructions. Therefore, we recommend providing additional details when fabricated information is found in LLM's output.
Furthermore, we found that using adjectives to describe expected outputs confuses LLM, leading to unexpected results. Avoiding such adjectives improves LLM comprehension of instructions.}

\noindent\textbf{Example in prompt might be harmful.}
% 在使用LLM时，有很多需求会导致我们希望在prompt中包含example，比如few-shot或者对于结果format的举例。然而，不当的例子会严重影响LLM的效率。
Including examples in prompts is sometimes necessary for few-shot learning or format demonstration. However, inappropriate examples can significantly reduce LLM efficiency.
% 比如，在APSR generation任务中，当我们尝试使用few-shot，并在prompt中包含不同API以及对应的APSR时，我们发现LLM往往在分析目标API时，只会考虑example中包含的APSR的类型，因此漏掉了很多其他类型的APSR。
For example, in the APSRs generation task, using a few-shot approach with specific APSRs examples caused the LLM to focus only on those types in the examples, missing other types. Therefore, it is crucial to ensure that examples do not limit the LLM's capabilities.

\noindent\textbf{Enhance prompt design with LLM.}
Leveraging LLM for prompt design can be highly beneficial. Providing the LLM with the intended purpose, input/output descriptions, and asking LLM to generate the prompt can help generate high-quality and well-structured task instructions.
% lesson of API usage
\subsection{Lessons for preventing API misuse}
\label{sec:discuss-misuse}
% 对API开发者和使用者的一些建议
Based on our analysis, we observe that error-prone or poorly documented documentations increase the likelihood of misuse. API developers should be mindful that users with varying experience may lack some common-sense knowledge and comprehensive security rule documentation is helpful. Additionally, careful error-checking in API calling code examples is crucial to provide accurate guidance to users.
\subsection{\change{Limitation and future work}}
\label{sec:discuss-limitation}
% 1. GPTAid依赖动态分析来分辨APSR是否正确。然而由于monitor的局限性，少部分APSR还是会被漏掉，比如：parameter must be unlock later，违反该APSR的代码并不会被GPTAid使用的monitor检测到异常。2. GPTAid的动态分析过程针对C/C++的用户态程序设计，因此GPTAid无法被直接使用到linux kernel或者其他语言的APSR生成上 3. 出于效率考虑，检测时仅仅使用intra分析会导致GPTAid漏掉部分bug。
\change{
For APSRs generation, \toolname{} relies on dynamic analysis to determine whether an APSR is correct. However, a common limitation of dynamic analysis is that monitors may fail to capture exceptions or detect bugs that don't cause exceptions, leading to missed APSRs. The monitors used by \toolname{} focus on memory issues and might miss a few APSRs, such as those related to logic errors (which do not cause exceptions) and lock-missing-unlock issues (whose exceptions cannot be captured).
Some existing studies~\cite{su2021logicbugs, zhang2024resolverfuzz} have extended the scope of dynamic analysis by enhancing capabilities of monitors. We plan to design more powerful monitors in the future to cover more APSRs. For bug detection, relying solely on intra-procedural analysis cause \toolname{} to miss some bugs, and we plan to improve the method of detection in future work.
% 除此之外，我们计划去探索软件中API误用数量的上限，以更好地分析API误用检测工作的效果并拓宽API误用检测的范围
Additionally, we plan to explore the upper limit on the number of API misuses in software to better assess the effectiveness of detection efforts and help detect more API misuses.
}
\section{Related Work}
\label{sec: related work}
\hit{
you should describe the other works briefly, including their method, results. In the end, you should summarize the difference between your work and their works.(if you cannot list the difference, do not mention this type of work.)
}
Recently, numerous approaches have emerged for generating APSRs and detecting API misuse, categorized as follows.

\noindent\textbf{Code-analysis based approaches. }
Some approaches generate APSRs by analyzing the source code of library APIs.
% goshawk使用NLP技术和数据流分析技术，通过分析API源码确定具有memory management功能的API以及对
% aurc
% analyse
% empirical
% hybrid
% limitition
Lyu et al.~\cite{Lyugoshawk2022} identifies functions related to memory management operations by analyzing the data flow of the source code and detects UAF and double free bugs.
Nguyen et al.~\cite{yenHybrid2023} use phrase-based Statistical machine translation (SMT) to translate code to the complete BE documentation.
Some approaches~\cite{ZhouAnalysing2017, Zhongempir2020} summarize some heuristic rules to extract parameter rules related to exception from API source code using static code analysis techniques.
Hu et al.~\cite{huaurc2023} utilizes static code analysis techniques to extract API rules about return value from library source code.\ignore{(treat information extract from library code as the correct rules).}
%不像这些方法，XX不受限于特殊的代码结构，XXX使用LLM分析library code来生成规则。
Unlike these approaches, \toolname{} utilizes LLM to analyze library code for APSR generation, without relying on specific code patterns.
Some approaches generate APSRs by analyzing client code that calls the library API.
% IPPO
% APISan,APEX
% Expose
% user guide Atrib。。
% limitation

Some approaches~\cite{yun2016apisan, kang2016apex} mine the correct usage patterns by identifying how most APIs are used.
Wen et al.~\cite{Wenexpose2019} mutate API usage in the client code according to predefined mutation rules and identify incorrect usage patterns by observing execution results to find API misuses.
% sinkfinder?
% 不像这些方法，受限于client code中有限的API并受client code中的错误影响会生成受限的错误的规则，XX使用LLM分析library code来生成规则。
Liu et al.~\cite{liuIPPO2021} detect bugs by checking the inconsistent security operations in a path-pair. 
Unlike these approaches, which are constrained by the limited APIs in the client code, resulting in the generation of restricted rules, \toolname{} employs LLM to analyze the library code and generate APSRs.
% 一些方法从test case的execution trace中挖掘FSA用以表示specification。
% deepm通过test case generation来获得多样化的execution trace并用以训练RNNLM来提取feature。通过对feature聚类挖掘出正确的FSA。adverm通过test case generation并生成反例来获取更多样化的execution trace并从中挖掘出的FSA。Unlike这些方法，\toolname{}不通过覆盖更多的execution trace来保证specification的完全性，而是通过分析源码来生成尽可能全面的规则，另外，\toolname{}生成的规则不局限于执行序列，还包括对参数值的细粒度约束。
Some approaches derive Finite State Automata (FSA) from test case execution traces to outline the specifications~\cite{le2018deepm, kang2021specm}. \delete{Le et al.~\cite{le2018deepm} generates diverse traces and use RNNLM to cluster features, mining correct FSAs. Kang et al.~\cite{kang2021specm} enhances trace diversity for FSA mining through test generation and counterexamples.}
Unlike these methods, \toolname{} analyzes API source code to formulate detailed rules, not limited to call sequence specifications but also detailing constraints on parameter values.

\noindent\textbf{Text-analysis based approaches. }
Some approaches use NLP techniques to generate APSRs from documentation.
% Advance   
% API graph
% Doc2Spec
% limitation
Lv et al.~\cite{lv2020rtfm} locate sentences with strong sentiment in documentations, and then extract API security rules by mining frequent patterns in these sentences.
Ren el al.~\cite{ren2020api} uses NLP techniques and heuristic algorithms to extract information from documentations, forming a fine-grained knowledge graph of API constraints.
% 不像这些方法，受限于文档的描述，只能从文档中提取有限的规则，XX通过分析library code生成更全面的规则。
Unlike these studies, which are limited by documentations and can only extract a limited number of rules, \toolname{} generates more rules by analyzing the library code.
% %%%%%%%%%%%%%%IGNORE%%%%%%%%%%%%%%%%%%%
\ignore{
\paragraph{Multi-source based.}
Some work has been done by combining two or more of the above sources of information.
% Analysing
% empire study
% AURC
% 这篇文章通过识别error handling的特殊结构从而识别出与error handling相关的参数条件，进而生成parameter specification.这篇文章将参数规则分成Nullness not allowed， Type restriction以及Range limitation这三类
Some work~\cite{ZhouAnalysing2017, Zhongempir2020} summarize some heuristic rules to extract parameter rules related to exception from documentation and API source code using NLP and code analysis techniques.
Hu et al.~\cite{huaurc2023} utilizes NLP and code analysis techniques to extract API rules about return value from documentation, library source code, and client code, and then detects documentation defects and API misuse regarding return values through cross-checking.
\yy{from the above paragraph, no results are listed and we cannot tell if this type of work is related to your work.}}
\ignore{
Nguyen et al.~\cite{yenHybrid2023} represent the source code of API and the behavioral exception (BE) documentation in the tree structure, and use phrase-based Statistical machine translation (SMT) to translate the complete BE documentation and the code by mapping and merging the clauses in the trees, so as to realize the goal of generating the BE documentation from the code.}
\section{Conclusion}
\label{sec:conclusion}
% 我们提出了XX，第一个使用LLM自动化生成API安全规则的工作。我们在4个库上应用了XX并针对部分API生成了XX条安全规则，填补了文档。并使用安全规则在XX个软件上找到了XX个bug。
We presented \toolname{}, the first work, to the best of our knowledge, to automatically generate APSRs by analyzing API source code with LLM and detect API misuse. 
\toolname{} utilizes an execution feedback-checking approach and code differential analysis to generate correct and concrete APSRs, and detect API misuse using APSRs.
On eight popular libraries, \toolname{} generated 579 APSRs, which were further investigated to enrich the documentation. Additionally, \toolname{} also found \change{210} unknown security bugs on 47 applications integrating these libraries which can lead to system crash and denial of services.
The result shows that \toolname{} is capable of protecting the safety use of APIs and it enlightens the future research work on exploiting LLMs for vulnerability detection.

\section*{Acknowledgements}
We would like to express our gratitude to our shepherd and reviewers for their valuable feedback, which greatly enhanced the quality of our paper.
The IIE authors are supported in part by CAS Project for Young Scientists in Basic Research (Grant No. YSBR-118), NSFC (92270204) and Youth Innovation Promotion Association CAS.

\bibliographystyle{IEEEtranS}
\bibliography{ref}

% \columnbreak

% \newpage
\vspace{20pt}
\section{Appendix}
\label{sec:appendix}
% 在这个部分，我们列出了对已知的API误用是否会导致能被捕获的runtime error的调查结果，见表xx。另外，我们列出了我们实验中使用的所有的库和软件的详细信息，见表xx。最后，我们列出了我们找到的所有的API误用的详细信息，见表xx。
% \newpage
\subsection{Detail Information}
In this section, we first present the results of our analysis on whether known API misuses trigger runtime errors detectable by monitoring tools, as shown in Table~\ref{tab:errmsg-misuse}. We also provide details of the libraries and applications used during evaluation in Table~\ref{tab:app_detail}. Finally, we present the details of all the API misuses identified by \toolname{}, as shown in Table~\ref{tab:bug-detail}.
% \vspace{10pt}
% \afterpage{\clearpage}
% \vfill
\begin{table}[htbp]
  \centering
  \caption{Study of known API misuse}
  \label{tab:errmsg-misuse}
  \begin{threeparttable}
  \resizebox{1\columnwidth}{!}{%
    \begin{tabular}{>{\centering\arraybackslash}p{4.7em}>{\centering\arraybackslash}p{4.6em}>{\centering\arraybackslash}p{6em}>{\centering\arraybackslash}p{5.4em}>{\centering\arraybackslash}p{2em}>{\centering\arraybackslash}p{2em}}
    \toprule
    Work  & CWE   & Impact & Catch? & \#Bugs & Total \\
    \midrule
    Advance & CWE-404 & DoS/Information Leakage & Sanitizer & 119   & \multirow{14}[2]{*}{749} \\
    Advance & CWE-690* & crash* & runtime error & 13    &  \\
    Goshawk & CWE-415* & \tabincell{c}{Memory\\Errors*} & Sanitizer & 45    &  \\
    Goshawk & CWE-416* & \tabincell{c}{Memory\\Errors*} & Sanitizer & 47    &  \\
    APP-Miner & CWE-476* & crash & runtime error & 89    &  \\
    APP-Miner & CWE-248* & crash & runtime error & 60    &  \\
    APP-Miner & CWE-404* & DoS   & Sanitizer & 4     &  \\
    APHP  & CWE-911 & DoS*  & Sanitizer & 222   &  \\
    APHP  & CWE-401 & DoS*  & Sanitizer & 61    &  \\
    APHP  & CWE-690 & crash* & runtime error & 6     &  \\
    APHP  & CWE-404 & DoS*  & Sanitizer & 83    &  \\
    \midrule
    Advance & \centering /     & malfunction & /     & 4     & \multirow{10}[2]{*}{47} \\
    Advance & CWE-253 & authentication errors & /     & 3     &  \\
    APP-Miner & CWE-414* & data modification & /     & 2     &  \\
    APP-Miner & CWE-190* & resource consumption & /     & 1     &  \\
    APP-Miner & CWE-563* & quality degradation & /     & 1     &  \\
    APHP  & CWE-235 & /     & /     & 36    &  \\
    \bottomrule
    \end{tabular}%
    }
    \begin{tablenotes}
    \footnotesize
    \item[1] / means we cannot determine this result.
    \item[2]  *  indicates results derived from our analysis of CWE and vulnerabilities,\\while absence of * indicates that the result is explicitly stated in the paper.
\end{tablenotes}
    \end{threeparttable}
  % \label{tab:addlabel}%
\end{table}%h

\begin{table}[h]
  \centering
  \small
  \caption{Libraries and Applications information}
  \resizebox{29em}{!}{
    \begin{tabular}{>{\centering\arraybackslash}m{1.41em}>{\centering\arraybackslash}m{3.92em}>{\centering\arraybackslash}m{1.63em}>{\centering\arraybackslash}m{15.04em}}
    \toprule
    \multicolumn{1}{c}{Library} & \multicolumn{1}{c}{Functionality} & \#API & \multicolumn{1}{c}{Applications} \\
    \midrule
    \multicolumn{1}{c}{libpcap} & \multicolumn{1}{c}{Network} & \multicolumn{1}{c}{71}    & masscan, SoftEtherVPN, nmap, john, ntopng, n2n, zmap, srs, tcpdump, freeradius-server \\
    \midrule
    \multicolumn{1}{c}{libxml2} & \multicolumn{1}{c}{XML parser} & \multicolumn{1}{c}{1614}  & openscap, php-src, aria2, collectd, postgres, vlc, ImageMagick, gpdb, ntopng, gdal \\
    \midrule
    \multicolumn{1}{c}{sqlite3} & \multicolumn{1}{c}{Database} & \multicolumn{1}{c}{294}   & netdata, php-src, leveldb, owntone-server, sqlcipher, ntopng, fluent-bit, gdal, wcdb, freeradius-server \\
    \midrule
    \multicolumn{1}{c}{openssl} &  \multicolumn{1}{c}{Cryptography} & \multicolumn{1}{c}{5478} & netdata, php-src, redis, curl, openssl, srs, SoftEtherVPN, nmap, fluent-bit, freeradius-server \\
    \midrule
    \multicolumn{1}{c}{libevent} & \multicolumn{1}{c}{Event handling} & \multicolumn{1}{c}{401}   & bitcoin-abc, evpp, owntone-server, seafile, transmission, libevent, gpdb, openvpn, kvrocks, bitcoin \\
    \midrule
    \multicolumn{1}{c}{libzip} & \multicolumn{1}{c}{\tabincell{c}{File Compression}} & \multicolumn{1}{c}{120}   & openrct2, hermes, ogre, monster-mash, radare2, rizin, xournalpp, idevicerestore, julius, cockatrice \\
    \midrule
    \multicolumn{1}{c}{zlib}  & \multicolumn{1}{c}{\tabincell{c}{Data Compression}} & \multicolumn{1}{c}{69}    & openrct2, netdata, imagemagick, curl, radare2, httpd, aria2, gpdb, postgres, kvrocks \\
    \midrule
    \multicolumn{1}{c}{libcurl} & \multicolumn{1}{c}{Network Transfer} & \multicolumn{1}{c}{76}    & curl, transmission, freeradius-server, gdal, openscap, gpdb, ntopng, collectd, fluent-bit, php-src \\
    \bottomrule
    \end{tabular}%
    }
  \label{tab:app_detail}%
\end{table}%

\clearpage

% Table generated by Excel2LaTeX from sheet 'Sheet1'

\clearpage

% test-3
\begin{table*}[t]
  \centering
  \caption{List of API misuse reported by \toolname{}}
  \label{tab:bug-detail}
  \begin{threeparttable}
  \makebox[\textwidth][c]{
  \begin{minipage}[t]{1\columnwidth}
  \vspace{0pt}
  \resizebox{1.05\columnwidth}{!}{
    \hspace{-6pt}
    \begin{tabular}{>{\centering\arraybackslash}m{2.11em}>{\centering\arraybackslash}m{3.02em}>{\centering\arraybackslash}m{11.93em}>{\centering\arraybackslash}m{2.94em}>{\centering\arraybackslash}m{1.85em}>{\centering\arraybackslash}m{1.56em}}
    \toprule
    \multicolumn{1}{c}{Library} & Software & {API} & {APSR} & Impact & {\#Bugs} \\
    \midrule
    \multicolumn{1}{c}{\multirow{6}[10]{*}{libpcap}} & \multirow{2}[2]{*}{tcpdump} & {pcap\_compile} & {$P_{2}$: $R_{1}$} & {Dos} & 2 \\
          & \multicolumn{1}{c}{} & {pcap\_findalldevs} & {$P_{1}$: $R_{1}$} & {Dos} & 1 \\
\cmidrule{2-6}          & zmap  & {pcap\_compile} & {$P_{2}$: $R_{1}$} & {Dos} & 1 \\
\cmidrule{2-6}          & ntopng & {pcap\_datalink} & {$P_{1}$: $R_{3}$} & {crash} & 1 \\
\cmidrule{2-6}          & n2n   & {pcap\_compile} & {$P_{2}$: $R_{1}$} & {Dos} & 1 \\
\cmidrule{2-6}          & nmap  & {pcap\_findalldevs} & {$P_{1}$: $R_{1}$} & {crash} & 1 \\
    \midrule
    \multicolumn{1}{c}{\multirow{2}[4]{*}{libxml2}} & php-src & \tabincell{c}{xmlParseURIReference} & {$P_{1}$: $R_{3}$} & {crash} & 3 \\
\cmidrule{2-6}          & openscap & \tabincell{c}{xmlXPathEval-\\-Expression} & {$P_{1}$: $R_{3}$} & {crash} & 2 \\
    \midrule
    \multicolumn{1}{c}{\multirow{7}[14]{*}{sqlite3}} & netdata & {sqlite3\_open} & {$P_{2}$: $R_{2}$} & {Dos} & 1 \\
\cmidrule{2-6}          & gdal  & {sqlite3\_open} & {$P_{2}$: $R_{2}$} & {Dos} & 1 \\
\cmidrule{2-6}          & fluent-bit & {sqlite3\_open} & {$P_{2}$: $R_{2}$} & {Dos} & 1 \\
\cmidrule{2-6}          & wcdb  & {sqlite3\_open\_v2} & {$P_{2}$: $R_{2}$} & {Dos} & 1 \\
\cmidrule{2-6}          & sqlcipher & {sqlite3\_open} & {$P_{2}$: $R_{2}$} & {Dos} & 1 \\
\cmidrule{2-6}          & freeradius-server & {sqlite3\_open\_v2} & {$P_{2}$: $R_{2}$} & {Dos} & 1 \\
\cmidrule{2-6}          & owntone-server & {sqlite3\_open} & {$P_{2}$: $R_{2}$} & {Dos} & 1 \\
    \midrule
    \multicolumn{1}{c}{\multirow{20}[18]{*}{openssl}} & \multirow{7}[2]{*}{nmap} & {EVP\_DigestInit} & {$P_{1}$: $R_{3}$} & {crash} & 1 \\
          & \multicolumn{1}{c}{} & \tabincell{c}{SSL\_CTX\_set\_cipher\_list} & {$P_{1}$: $R_{3}$} & {crash} & 1 \\
          & \multicolumn{1}{c}{} & \tabincell{c}{EVP\_CIPHER\_CTX\_set\\\_padding} & {$P_{1}$: $R_{3}$} & {crash} & 2 \\
          & \multicolumn{1}{c}{} & {BN\_bin2bn} & {$P_{1}$: $R_{3}$} & {crash} & 1 \\
          & \multicolumn{1}{c}{} & {EVP\_EncryptInit\_ex} & {$P_{1}$: $R_{3}$} & {crash} & 1 \\
          & \multicolumn{1}{c}{} & {EVP\_DecryptInit\_ex} & {$P_{1}$: $R_{3}$} & {crash} & 1 \\
          & \multicolumn{1}{c}{} & {EVP\_DigestInit\_ex} & {$P_{1}$: $R_{3}$} & {crash} & 2 \\
\cmidrule{2-6}          & \multirow{6}[2]{*}{\tabincell{c}{freeradius-\\server}} & {EVP\_DigestSignInit} & {$P_{1}$: $R_{3}$} & {crash} & 2 \\
          & \multicolumn{1}{c}{} & \tabincell{c}{X509\_STORE\_CTX\_set\\\_ex\_data} & {$P_{1}$: $R_{3}$} & {crash} & 1 \\
          & \multicolumn{1}{c}{} & {EVP\_DecryptInit\_ex} & {$P_{1}$: $R_{3}$} & {crash} & 2 \\
          & \multicolumn{1}{c}{} & {EVP\_EncryptInit\_ex} & {$P_{1}$: $R_{3}$} & {crash} & 2 \\
          & \multicolumn{1}{c}{} & \tabincell{c}{EVP\_CIPHER\_CTX\_set\\\_key\_length} & {$P_{1}$: $R_{3}$} & {crash} & 1 \\
          & \multicolumn{1}{c}{} & {EVP\_DigestInit\_ex} & {$P_{1}$: $R_{3}$} & {crash} & 3 \\
\cmidrule{2-6}          & fluent-bit & {PEM\_read\_bio\_PrivateKey} & {$P_{1}$: $R_{3}$} & {crash} & 1 \\
\cmidrule{2-6}          & php-src & {EVP\_DigestInit} & {$P_{1}$: $R_{3}$} & {crash} & 2 \\ 
\cmidrule{2-6}          & \multirow{2}[2]{*}{openssl} & {SSL\_ctrl} & {$P_{1}$: $R_{3}$} & {crash} & 3 \\
          & \multicolumn{1}{c}{} & {EVP\_DigestInit\_ex} & {$P_{1}$: $R_{3}$} & {crash} & 2 \\
\cmidrule{2-6}          & srs   & {SSL\_CTX\_set\_cipher\_list} & {$P_{1}$: $R_{3}$} & {crash} & 1 \\
    \midrule
    \end{tabular}
    }
    \end{minipage}
    % \columnbreak
\begin{minipage}[t]{1.05\columnwidth}
\vspace{0pt}
\resizebox{1.05\columnwidth}{!}{
    \hspace{6pt}
    \begin{tabular}{>{\centering\arraybackslash}m{2.11em}>{\centering\arraybackslash}m{3.02em}>{\centering\arraybackslash}m{11.93em}>{\centering\arraybackslash}m{2.94em}>{\centering\arraybackslash}m{1.85em}>{\centering\arraybackslash}m{1.56em}}
    \toprule
    \multicolumn{1}{c}{Library} & Software & {API} & {APSR} & {Impact} & {\#Bugs} \\
    \midrule
    \multicolumn{1}{c}{\multirow{12}[6]{*}{openssl}} & \multirow{8}[2]{*}{\tabincell{c}{SoftEther-\\-VPN}} & {EVP\_DigestInit\_ex} & {$P_{1}$: $R_{3}$} & {crash} & 1 \\
          & \multicolumn{1}{c}{} & {DH\_set0\_pqg} & {$P_{1}$: $R_{3}$} & {crash} & 1 \\
          & \multicolumn{1}{c}{} & {BN\_bn2bin} & {$P_{2}$: $R_{3}$} & {crash} & 1 \\
          & \multicolumn{1}{c}{} & {SSL\_CTX\_set\_verify} & {$P_{1}$: $R_{3}$} & {crash} & 1 \\
          & \multicolumn{1}{c}{} & {SSL\_set\_ex\_data} & {$P_{1}$: $R_{3}$} & {crash} & 1 \\
          & \multicolumn{1}{c}{} & {SSL\_set\_fd} & {$P_{1}$: $R_{3}$} & {crash} & 1 \\
          & \multicolumn{1}{c}{} & {SSL\_CTX\_set\_options} & {$P_{1}$: $R_{3}$} & {crash} & 7 \\
          & \multicolumn{1}{c}{} & \tabincell{c}{SSL\_CTX\_set\_ssl\\\_version} & {$P_{1}$: $R_{3}$} & {crash} & 3 \\
\cmidrule{2-6}          & \multirow{3}[2]{*}{netdata} & {EVP\_DigestInit\_ex} & {$P_{1}$: $R_{3}$} & {crash} & 2 \\
          & \multicolumn{1}{c}{} & {SSL\_CTX\_get\_options} & {$P_{1}$: $R_{3}$} & {crash} & 1 \\
          & \multicolumn{1}{c}{} & {SSL\_CTX\_set\_options} & {$P_{1}$: $R_{3}$} & {crash} & 1 \\
\cmidrule{2-6}          & \multirow{2}[2]{*}{redis} & {ERR\_error\_string\_n} & {$P_{2}$: $R_{3}$} & {crash} & 1 \\
          & \multicolumn{1}{c}{} & {SSL\_CTX\_set\_options} & {$P_{1}$: $R_{3}$} & {crash} & 1 \\
          \midrule
    \multicolumn{1}{c}{\multirow{29}[12]{*}{libevent}} & \multirow{4}[2]{*}{kvrocks} & {evdns\_base\_resolv\_conf\_parse} & {$P_{1}$: $R_{3}$} & {crash} & 1 \\
          & \multicolumn{1}{c}{} & {evdns\_base\_set\_option} & {$P_{1}$: $R_{3}$} & {crash} & 1 \\
          & \multicolumn{1}{c}{} & {evdns\_base\_nameserver\_ip\_add} & {$P_{1}$: $R_{3}$} & {crash} & 1 \\
          & \multicolumn{1}{c}{} & {evdns\_base\_nameserver\_add} & {$P_{1}$: $R_{3}$} & {crash} & 1 \\
\cmidrule{2-6}          & \multirow{5}[2]{*}{evpp} & {event\_add} & {$P_{1}$: $R_{3}$} & {crash} & 2 \\
          & \multicolumn{1}{c}{} & {evhttp\_connection\_base\_new} & {$P_{1}$: $R_{3}$} & {crash} & 3 \\
          & \multicolumn{1}{c}{} & {evhttp\_make\_request} & {$P_{1}$: $R_{3}$} & {crash} & 2 \\
          & \multicolumn{1}{c}{} & {evhttp\_uri\_get\_path} & {$P_{1}$: $R_{3}$} & {crash} & 1 \\
          & \multicolumn{1}{c}{} & {bufferevent\_setcb} & {$P_{1}$: $R_{3}$} & {crash} & 2 \\
\cmidrule{2-6}          & \multirow{3}[2]{*}{\tabincell{c}{owntone-\\server}} & {event\_add} & {$P_{1}$: $R_{3}$} & {crash} & 4 \\
          & \multicolumn{1}{c}{} & {bufferevent\_free} & {$P_{1}$: $R_{3}$} & {crash} & 1 \\
          & \multicolumn{1}{c}{} & {bufferevent\_setcb} & {$P_{1}$: $R_{3}$} & {crash} & 1 \\
\cmidrule{2-6}          & seafile & {event\_add} & {$P_{1}$: $R_{3}$} & {crash} & 2 \\
\cmidrule{2-6}          & \tabincell{c}{transmi-\\-ssion} & {evhttp\_set\_allowed\_methods} & {$P_{1}$: $R_{3}$} & {crash} & 1 \\
\cmidrule{2-6}          & \multirow{15}[2]{*}{libevent} & {bufferevent\_setcb} & {$P_{1}$: $R_{3}$} & {crash} & 32 \\
          & \multicolumn{1}{c}{} & {bufferevent\_getfd} & {$P_{1}$: $R_{3}$} & {crash} & 1 \\
          & \multicolumn{1}{c}{} & {bufferevent\_pair\_get\_partner} & {$P_{1}$: $R_{3}$} & {crash} & 1 \\
          & \multicolumn{1}{c}{} & {bufferevent\_enable} & {$P_{1}$: $R_{3}$} & {crash} & 4 \\
          & \multicolumn{1}{c}{} & {bufferevent\_get\_input} & {$P_{1}$: $R_{3}$} & {crash} & 1 \\
          & \multicolumn{1}{c}{} & {bufferevent\_get\_output} & {$P_{1}$: $R_{3}$} & {crash} & 2 \\
          & \multicolumn{1}{c}{} & {evhttp\_connection\_set\_timeout} & {$P_{1}$: $R_{3}$} & {crash} & 1 \\
          & \multicolumn{1}{c}{} & \tabincell{c}{evdns\_base\_nameserver\\\_ip\_add} & {$P_{1}$: $R_{3}$} & {crash} & 17 \\
          & \multicolumn{1}{c}{} & {evdns\_base\_resolv\_conf\_parse} & {$P_{1}$: $R_{3}$} & {crash} & 2 \\
          & \multicolumn{1}{c}{} & {event\_base\_dispatch} & {$P_{1}$: $R_{3}$} & {crash} & 3 \\
          & \multicolumn{1}{c}{} & {evdns\_base\_set\_option} & {$P_{1}$: $R_{3}$} & {crash} & 2 \\
          & \multicolumn{1}{c}{} & {event\_add} & {$P_{1}$: $R_{3}$} & {crash} & 48 \\
          & \multicolumn{1}{c}{} & {evhttp\_make\_request} & {$P_{1}$: $R_{3}$} & {crash} & 1 \\
          & \multicolumn{1}{c}{} & {evdns\_base\_nameserver\_add} & {$P_{1}$: $R_{3}$} & {crash} & 1 \\
          & \multicolumn{1}{c}{} & {bufferevent\_setwatermark} & {$P_{1}$: $R_{3}$} & {crash} & 1 \\
    \midrule
    \end{tabular}%
    }
    \end{minipage}
    }
    \begin{tablenotes}
    \footnotesize
    \item[1] APSR are abridged to improve clarity and fit within the table.
    \item[2] $P_{n}$ denotes Parameter $n$.
    \item[3] $R_{1}$ indicates the rule: must be freed later.
    \item[4] $R_{2}$ indicates the rule: must be closed later.
    \item[5] $R_{3}$ indicates the rule: must not be NULL.
\end{tablenotes}
  \label{tab:addlabel}%
  \end{threeparttable}
\end{table*}%

\clearpage
\subsection{LLM Strategies Selection} 
\balance
\label{sec:temp-selection}
\begin{figure}[htbp]
    \centering
    \includegraphics[width=1\linewidth]{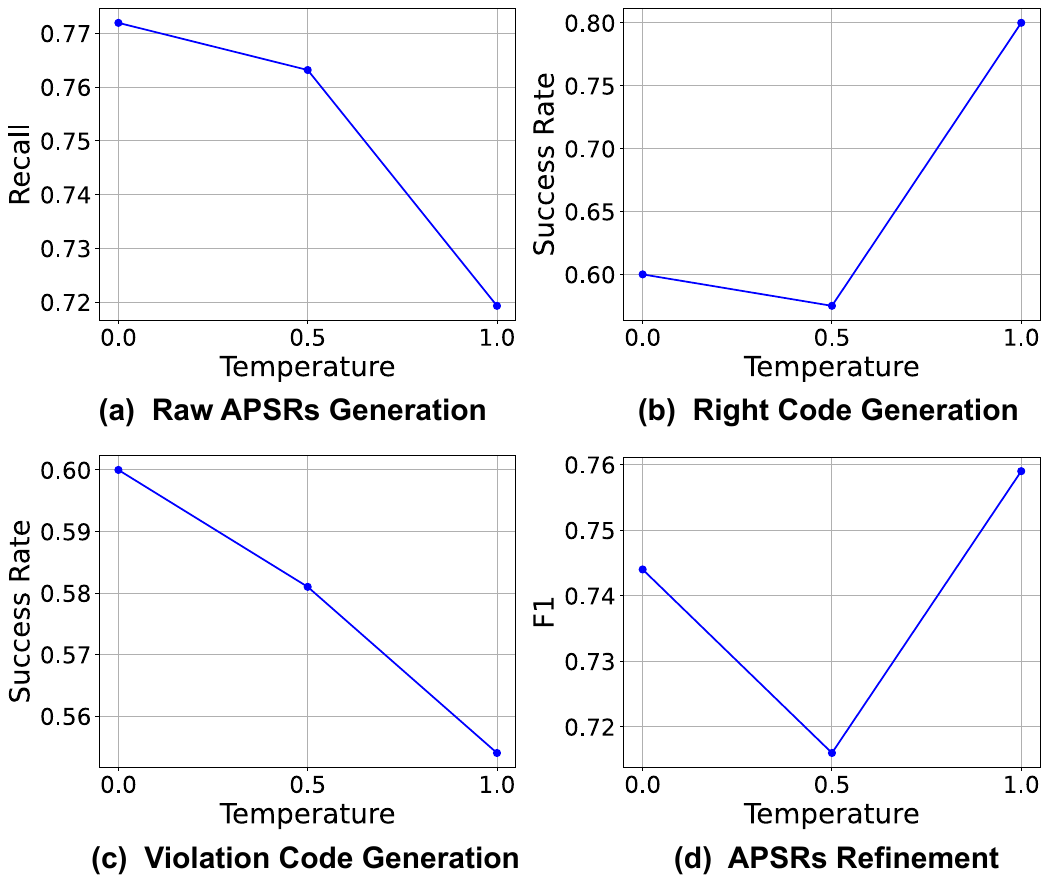}
    \caption{Results of Temperature selection}
    \label{fig:temp-info}
    % \vspace{10pt}
\end{figure}
\vspace {3pt}\noindent$\bullet$\space\textit{Ground-Truth dataset for LLM Strategies Selection ($D_{sgt}$).}
To select the temperature of model,  we randomly selected 40 APIs from four libraries that differ from the APIs in $D_{gt}$. We then manually generated 134 APSRs by analyzing the API source code and documentation.
% 考虑到temperature会影响到结果的准确性与稳定性，我们在0-2的temperature区间上使用不同的temperature对toolname在GT上的效果进行了评估。经过多次实验我们发现当temperature大于1时，会出现结果输出乱码的现象，因此我们在0-1的区间上使用temperature 0，0.5，1进行了三组不同temperature的实验。
% 针对四个使用LLM的模块：Raw APSRs Generation, Right Code Generation, Violation Code Generation, APSRs refinement, 我们分别在Dgt上对其效果进行了评估。针对Raw APSRs Generation模块，我们将Raw APSR recall作为评估指标以确定能生成最多Raw APSR的temperature；针对Right Code Generation和Violation Code Generation模块，我们将生成代码的成功率作为评估指标， 针对APSRs refinement模块，我们将生成的APSR的F1作为评估指标。实验结果如图XX所示。当四个模块temperature分别为xx,xx,xx,xx时，得到最好的结果，因此我们使用该组temperature作为toolname使用的temperature。

\paragraph{Prompt design. }In this section, we assessed how different prompting methods (zero-shot, few-shot, Chain-Of-Thought) affect raw APSR generation on $D_{sgt}$. The recall rates were 48.20\% for zero-shot, 65.80\% for few-shot, and 53.50\% for Chain-Of-Thought. Although few-shot had the highest recall, it mostly created APSRs types similar to those in the prompts and ignored others, which could potentially limit LLM's functionality. To foster diverse APSRs generation, we used the Chain-Of-Thought method for prompt design.
% 经过我们的分析，zero-shot, few-shot, Chain-Of-Thought的recall分别为：xxx。虽然few-show的prompt得到了最高的recall，然而经过分析我们发现，该方法得到的结果更倾向于生成prompt中涉及到的规则类型，会漏掉绝大部分其他类型的规则，这会使LLM的功能受限。为了LLM能够生成多样化的规则，我们使用了Chain-Of-Thought方法来设计prompt。

\paragraph{Temperature selection. }In this section, we evaluated the effectiveness of \toolname{} on $D_{sgt}$ using different temperatures between 0 and 2. After several experiments, we found that setting the temperature above 1 results in meaningless, messy output. Therefore, we only conducted experiments with three temperatures: 0, 0.5, and 1. For each of the four takss using LLMs: Raw APSRs Generation, Right Code Generation, Violation Code Generation, APSRs refinement, we evaluated their effectiveness on $D_{sgt}$. 
%我们的目的是使LLM生成尽可能多的raw APSR，因此 
% 我们希望LLM按照指定的要求生成正确的代码，
% 我们希望对APSR的refine能使APSR尽可能正确且全面，因此我们需要综合考察precision和recall这两个指标，因此
For the Raw APSRs Generation, our goal is to maximize the number of APSRs, so we employ Raw APSR recall as an evaluation metric to identify the temperature generating the maximum number of Raw APSRs. For Right Code Generation and Violation Code Generation, we aim for LLM to generate accurate code in accordance with the specified requirements, so the success rate of code generation serves as the evaluation metric. In the APSR Refinement, our goal is to refine the APSRs to make them concrete. We focus on two key metrics: precision and recall. To assess the performance, we use the F1 score of the generated APSRs. The experimental results are shown in Figure~\ref{fig:temp-info}. The best performance is obtained when the temperatures for the four tasks are 0,1,0,1 respectively, which is the temperature used by \toolname{}.

\end{document}